\documentclass[%
 reprint,
showpacs,preprintnumbers,
 amsmath,amssymb,
 aps,
pra,
floatfix,
]{revtex4-1}

\usepackage[draft]{hyperref}
\usepackage{amsmath}
\usepackage{empheq}  
\usepackage{graphicx}	
\usepackage{epstopdf}
\usepackage{ctable}
\usepackage{xcolor}
\usepackage{epsfig}
\usepackage{dcolumn}
\usepackage{bm}

\begin{document}
\title{Triply-resonant micro-optical parametric oscillators based on Kerr nonlinearity: nonlinear loss, unequal resonance-port couplings, and coupled-cavity implementations}


\author{Xiaoge Zeng}
\author{Milo\v{s} A. Popovi\'{c}}
\email{Correspondence by email to milos.popovic@colorado.edu}
\affiliation{Department of Electrical, Computer and Energy Engineering and Department of Physics, University of Colorado, Boulder, CO 80309, USA}    
\date{\today}

\begin{abstract}

\noindent We develop a theoretical model of triply-resonant optical parametric oscillators (OPOs) based on degenerate four-wave mixing (FWM) that includes physics and degrees of freedom relevant to microphotonic (on-chip) device implementations, including nonlinear loss, a general resonant mode field structure, and mode-selective coupling to external ports.  The coupled mode theory (CMT) model addresses the effect of two-photon absorption (TPA) and free-carrier absorption (FCA) on parametric gain and oscillation thresholds, and ultimately on the optimum design for an OPO.  The model goes beyond a typical free-space cavity configuration by incorporating a full modal analysis that admits distributed modes with non-uniform field distribution, relevant to photonic microcavity systems on chip.  This leads to a generalization of the concept of nonlinear figure of merit (NFOM) to a vector of coefficients.  In addition, by considering unconstrained signal, pump and idler resonance coupling strengths to excitation ports, not usually available in simple cavity geometries, we show that the efficiency-maximizing design will have unequal external Q for the three resonances.  We arrive at generalized formulas for OPO oscillation threshold that include nonlinear absorption and free carrier lifetime, and provide a normalized solution to the design problem in the presence of nonlinear loss in terms of optimum choice of coupling.  Based on the results, we suggest a family of coupled-cavity systems to implement optimum four-wave mixing, where control of resonant wavelengths can be separated from optimizing nonlinear conversion efficiency, and where furthermore pump, signal, and idler coupling to bus waveguides can be controlled independently, using interferometric cavity supermode coupling as an example.  Using the generalized NFOM, we address the efficiency of single and multi-cavity geometries, as well as standing and traveling wave excitations.

\end{abstract}
\maketitle

\section{Introduction} \label{sec:introduction}
\noindent On-chip, coherent light generation is of interest
for many classical photonics applications, including light sources at wavelengths where gain media are underdeveloped, optical frequency comb generation \cite{del2007optical, foster2011silicon}, and optical data stream wavelength conversion \cite{turner2008ultra}.  It is also of interest in quantum optics, for heralded single photon \cite{davancco2012telecommunications} and correlated photon pair generation\cite{engin2012photon, azzini2012ultra}.  One promising approach employs microcavity-enhanced optical parametric gain based on third-order nonlinearity\cite{kippenberg2004kerr}, $\chi^{(3)}$.  The nonlinearity is strong in semiconductors like silicon as well as a number of nonlinear glasses, and can be greatly enhanced in a microcavity due to strong transverse spatial confinement and large effective interaction length \cite{Turner:08}.  With high enough parametric gain, optical parametric oscillation (OPO) is possible, where a single input pump wavelength enables oscillation at two other wavelengths resonant in the microcavity, which substantial energy conversion.  Previous demonstrations include optical parametric oscillation based on four-wave mixing (FWM) in silica microtoroids \cite{kippenberg2004kerr}, silica \cite{razzari2009cmos} and silicon nitride (Si$_3$N$_4$) microring resonators\cite{levy2009cmos}, where nonlinear loss due to two-photon-absorption (TPA) is ignorable.  It is of interest to investigate the fundamental limits of micro-OPO performance, and find conditions of maximum conversion efficiency for given material parameters. This is important because on-chip microphotonic cavity geometries have access to greater degrees of freedom in design than either bulk optics or thin film optics \cite{diederichs2006parametric}, as previously shown in optimal filters \cite{popovic2007sharply,popovic2008experimental}, modulators \cite{Sacher:08,popovic2010resonant}, light trapping \cite{yanik2004stopping,dahlem2010dynamical}.  Also, a first-principles look is important because on-chip implementations in semiconductors may have substantial nonlinear losses in addition to linear loss, including two-photon absorption (TPA), and TPA-induced free-carrier absorption \cite{dimitropoulos2005lifetime,rong2005continuous}.  
Finally, even the typical scaling of resonator quality factor due to the linear losses -- the linear unloaded or loss Q -- is different in e.g. tabletop cavities, where it is normally limited by lumped mirror loss, and integrated microring or photonic crystal cavities, where it is normally dominated by a distributed loss (per unit length) that may be due to waveguide surface roughness, material absorption, or bulk scattering.  This has an impact on scaling of designs.
\begin{figure}[b]
\vspace{-16pt}
	\includegraphics[width=3.25in]{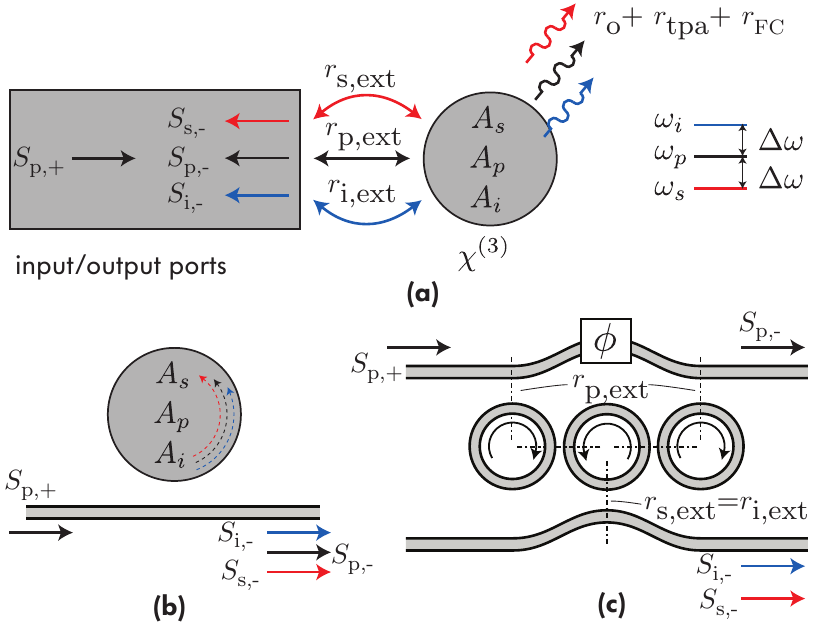}
\vspace{-6pt}
	\caption{(a) Illustration of the micro-OPO model including a multimode resonator; 
	(b) a traveling-wave resonant structure enables separated input and output ports; 
	(c) example proposed multimode resonator based on 3 coupled microring cavities, showing an approach to unequal pump and signal/idler external coupling \cite{zeng2013optimum}.\label{fig1_schematics}}
\vspace{-3pt}
\end{figure}

In this paper, we present a general analysis of OPOs based on degenerate-pump FWM, which leads to a design approach enabling maximum possible conversion efficiency.  Previously, temporal coupled-mode theory (CMT) has been used successfully to analyze resonant nonlinear systems \cite{haus1984waves}, including ones with four-wave mixing \cite{Ramirez2011Degenerate}.  We develop a temporal coupled mode theory (CMT) model of the parametric oscillator in Fig.~\ref{fig1_schematics}, and normalize it with respect to linear losses, giving a very general representation of the OPO design problem, in terms of normalized pump power and the material nonlinear figure of merit (NFOM).  We also address free carrier losses by a normalized parameter.  The optimum design solutions to this model are thus broadly applicable across a wide parameter space, and we are able to draw general conclusions about oscillation thresholds, conversion efficiency, and optimum coupling independent of the particular device geometry.

Our model bears out some fundamental limitations of conventional (including integrated photonic) OPO implementations, and suggests new, more general geometries that are capable of realizing these optimum OPO designs.  In particular, different external-coupling linewidths are desirable for the pump, signal and idler wavelength resonances; and, dispersion engineering can be accomplished in part using interferometric cavity and coupled-cavity configurations in addition to design of the waveguide cross-section.  These amount to resonant ``photonic molecules'' that allow effective engineering of both resonance frequencies (states) and linewidths (lifetimes) independently.

The study also leads to a few general conclusions.  First, we show that there is a critical value of the nonlinear figure of merit in materials beyond which parametric oscillation is not possible in a certain cavity mode structure.  On the other hand, we show, for example, that crystalline silicon structures can oscillate in principle, even in the telecom band where nonlinear losses are present.  Second, we provide a general oscillation threshold formula for OPOs that includes both two-photon absorption (TPA) and free-carrier absorption (FCA) losses in a normalized way.  Third, in view of optical confinement issues both in basic cavities and the more complex ``photonic molecule'' geometries, we devote a section to comparing single cavity to multicavity geometries, as well as traveling wave to standing wave mode excitation.  The latter are both possible excitations in a traveling wave cavity like a microring resonator, while only standing wave excitation is possible in a standing-wave single-mode cavity like a photonic crystal microcavity.
In a simple photonic structure that contains a single nonlinear material (a ring with traveling-wave mode where a single material's nonlinearity dominates), the nonlinear figure of merit (NFOM) is unchanged from the bulk material value by overlap integrals.  On the other hand, in both standing vs. traveling wave excitation, and single vs. multiple cavity geometries, we find that the parametric gain and nonlinear losses differ, and, furthermore, even different types of nonlinear loss (e.g. absorption of two pump photons vs. absorption of a signal and an idler photon) differ.  This means that a single NFOM is no longer sufficient to describe performance.  In the simplest cases, a single \emph{effective} NFOM can be defined, while in more complex geometries, the NFOM concept needs to be generalized.  We supplement the NFOM with a vector of coefficients, that we call $\vec{d}$ ($d_\mathrm{mn}$), that reflect the mode topology contribution.  The material NFOM together with the vector $\vec{d}$ can then be thought of as a generalized NFOM and used to evaluate various designs.

We also provide some practical examples of experimental relevance, to give the reader some orientation. For example, $0.21\,$mW of pump light at $1.55\,\mu$m in silicon microrings (R=$3\,\mu$m) with intrinsic quality factor ($Q_o$) of $10^6$ and a free-carrier-lifetime of $60\,$ps can produce parametric oscillation with about $0.1\%$ conversion efficiency (free-carrier-lifetime of $12.2\,$ps has been demonstrated\cite{turner2010ultrashort}), and the efficiency can reach $2\%$ as free-carrier-lifetime approaches zero (e.g. via active carrier sweepout \cite{dimitropoulos2005lifetime, rong2005continuous, engin2012photon}). In general, an OPO based on silicon as the source of parametric gain cannot produce oscillation with conversion efficiencies on the order of 1 at $1550\,$nm due to TPA, but we show that they can achieve oscillation at up to a few percent conversion in principle. At wavelengths above around $2.2\,\mu$m (i.e.\,below the half bandgap energy of silicon), efficiency close to $17\%$ is achievable in a microring (R=$7\,\mu$m) with a pump power of $1\,$mW and a cavity intrinsic Q of $10^6$.  In Si$_3$N$_4$ microrings (R=$15\,\mu$m), a weaker nonlinearity but absence of TPA in the telecom band enables conversion efficiencies approaching $10\%$ at about $9\,$mW pump power with a $Q_o$ of $10^6$, where the maximum possible is $50\%$ to each of the signal and idler (details in Section~\ref{sec:ExampleDesigns} and Table~\ref{table:table1}).

The paper is organized as follows.  In Section~\ref{sec:physical_model}, we provide our CMT model of the general parametric oscillator, and discuss approximations and assumptions.  In Section~\ref{sec:normalized_model}, we normalize the model with respect to linear and nonlinear loss, so that the results of computations are universally applicable across various microcavity systems.  In Section~\ref{sec:this_approach}, we explain the approach for finding the optimum design, and in Sections~\ref{sec:pumptpaonly}--\ref{sec:fulltpaandfca} we solve the problem of optimum OPO design approximately and exactly, and provide universal design curves for situations with and without TPA and FCA losses.  In Section~\ref{sec:efffom}, we generalize the resonator design to higher order systems including ``photonic molecules'' that allow systematic state engineering and state-selective coupling engineering, and compare standing-wave and traveling-wave excitations.  In this context, we discuss OPO architectures that offer certain advantages over single-cavity designs.

\section{Physical model of a parametric oscillator} \label{sec:physical_model}
We begin by constructing a coupled mode theory in time (CMT) model \cite{haus1984waves, Rodriguez2007chi2}, valid to describe the dynamics of linear and nonlinear phenomena in microcavities in the weak nonlinearity regime, $\chi^{(3)}|E|^2\ll 1$ (valid in practical OPOs).  We consider only three resonantly-enhanced, interacting frequencies (one signal, a degenerate pump and one idler), which is a valid assumption if the optical resonator is dispersion engineered to allow phase matching to the pump for only one pair of output signal and idler wavelengths.  This is unlike a periodic comb generator \cite{kippenberg2004kerr} 
which produces many output wavelengths, but can be accomplished with appropriate resonator and dispersion engineering \cite{atabaki2011ultra} and, we expect, is the optimal way to use the parametric gain when the objective is efficient generation of a single pair of signal/idler output wavelengths.  We focus on this case here because it is the simplest, but much of the intuition provided by our conclusions will apply more broadly.

The CMT model for the three-resonance system illustrated in Fig.~\ref{fig1_schematics}, assuming frequency (energy) matching, is
\begin{subequations}
\begin{align}
\frac{dA_s}{dt} & = -r_\mathrm{s,tot} A_s -j\omega_s\beta_\mathrm{fwm,s} A_p^2 A_i^{\ast}\label{eqn:dtAs}\\
\frac{dA_p}{dt} & = -r_\mathrm{p,tot} A_p -2j\omega_p\beta_\mathrm{fwm,p} A_p^{\ast}A_s A_i - j \sqrt{2r_\mathrm{p, ext}}\,S_{p,+} \label{eqn:ap dynamics}\\
\frac{dA_i}{dt} & = -r_\mathrm{i,tot} A_i -j\omega_i\beta_\mathrm{fwm,i} A_p^2 A_s^{\ast}\label{eqn:dtAi}\\
S_{s,-} & = - j \sqrt{2 r_\mathrm{s,ext}} A_s\\
S_{p,-} & = S_{p,+} - j \sqrt{2 r_\mathrm{p,ext}} A_p\\
S_{i,-} & = - j \sqrt{2 r_\mathrm{i,ext}} A_i \label{eqn:dtSim}
\end{align}
\end{subequations}
where $A_k(t)$, $k \in \{p,s,i\}$, are the cavity energy-amplitude \emph{envelopes} for light at pump, signal and idler frequencies; $S_{k,+}\,(S_{k,-})$ is the power-amplitude \emph{envelope} in the input (output) port for each resonant mode; $\omega_k$ are the angular frequencies of the interacting modes; and $\beta_\mathrm{fwm,k}$ are the FWM (parametric gain) coefficients, related to modal field overlap integrals in Appendix~\ref{sec:app:ovlintegrals}.  By ``envelope'', we mean that $A_k(t)$ is related to the usual CMT amplitude\cite{haus1984waves} $a_k(t)$ by $a_k(t) \equiv A_k(t) e^{j \omega_k t}$.  For simplicity, we normalize mode field patterns to unity energy or power, such that $|A_k|^2$ is the energy of resonant mode $k$ and $|S_{k,+}|^2$ ($|S_{k,-}|^2$) is the inbound (outbound) power in guided mode $k$.  This corresponds to each term in the denominators of overlap integrals (\ref{eqn:beta_fwm}) and (\ref{eqn:beta_tpa}) in Appendix~\ref{sec:app:ovlintegrals} being set to unity.  The above equations are valid for the case of resonance frequency (energy) matching, $2\omega_p=\omega_s+\omega_i$, and don't contain the detuning $\exp [j(2\omega_p-\omega_s-\omega_i)]$ terms, because we focus on the most efficient case.  From these CMT equations one can derive the energy conservation law $2\omega_p\beta_\mathrm{fwm,p}^\ast = \omega_s\beta_\mathrm{fwm,s}+ \omega_i\beta_\mathrm{fwm,i}$. As discussed in Appendix~\ref{sec:app:ovlintegrals}, the coefficients $\beta_\mathrm{fwm,s}$, $\beta_\mathrm{fwm,p}^\ast$, and $\beta_\mathrm{fwm,i}$ are identical except for the tensor element of $\chi^{(3)}$ that they contain.  Under the assumption of full permutation symmetry\cite{boyd2008nonlinear}, these tensor elements and hence the foregoing coefficients, are equal.  For the purposes of the remainder of this paper, we define a single $\beta_\mathrm{fwm}$, and define
\begin{align}
\beta_\mathrm{fwm,s}= \beta_\mathrm{fwm,p}^\ast= \beta_\mathrm{fwm,i} \equiv \beta_\mathrm{fwm}.
\end{align}

Decay rate $r_\mathrm{k,tot}$ is the total energy amplitude decay rate for mode $k$ (due to both loss and coupling to external ports), where \cite{fishman2011sensitive}
\begin{align}
r_\mathrm{s,tot} = \,&r_\mathrm{s,o}+ r_\mathrm{s,ext} +r_\mathrm{FC}
+ \omega_s\left(\beta_\mathrm{tpa,ss}|A_s|^2 + \right .\nonumber\\
& \left . 2\beta_\mathrm{tpa,sp}|A_p|^2 + 2\beta_\mathrm{tpa,si}|A_i|^2\right) \nonumber\\
r_\mathrm{p,tot} = \,&r_\mathrm{p,o}+ r_\mathrm{p,ext} +r_\mathrm{FC}
+ \omega_p\left(2\beta_\mathrm{tpa,sp}|A_s|^2 + \right .\nonumber\\
& \left . \beta_\mathrm{tpa,pp}|A_p|^2 + 2\beta_\mathrm{tpa,ip}|A_i|^2\right) \nonumber\\
r_\mathrm{i,tot} = \,&r_\mathrm{i,o}+ r_\mathrm{i,ext} +r_\mathrm{FC}
+ \omega_i\left(2\beta_\mathrm{tpa,si}|A_s|^2 + \right .\nonumber\\
& \left . 2\beta_\mathrm{tpa,ip}|A_p|^2 + \beta_\mathrm{tpa,ii}|A_i|^2\right).\label{eqn:loss1}
\end{align}
Here, $r_\mathrm{k,o}$, $k \in \{s,p,i\}$ is the linear loss rate of mode $k$, $r_\mathrm{k,ext}$ is the coupling rate to an external port (e.g. waveguide, see Fig.~\ref{fig1_schematics}) and $\beta_\mathrm{tpa,mn}$ is the two-photon absorption coefficient due to absorption of a photon each from modes $m$ and $n$ ($m,n \in \{s,p,i\}$). $\beta_\mathrm{tpa,mn}$ should not be confused with the coefficient $\beta_\mathrm{TPA}$ typically used in the nonlinear optics literature, which is a bulk (plane wave) value, is defined through $dI/dz = -\beta_\mathrm{TPA}I^2$ and represents `nonlinear loss' per unit length; $\beta_\mathrm{tpa,mn}$ here has units of `nonlinear loss' per unit time (for a resonant mode) and includes a spatial mode overlap integral to account for the spatial inhomogeneity of the field and lump it into a single effective factor (defined in Appendix~\ref{sec:app:ovlintegrals}).

Finally, the decay rate includes a contribution due to free-carrier absorption (FCA).  The FCA loss rate, $r_\mathrm{FC}$, is not a constant like the other rates and coefficients $r_\mathrm{k,o}$, $r_\mathrm{k,ext}$ and $\beta_\mathrm{tpa,mn}$ in Eq.~(\ref{eqn:loss1}), but depends on intensities.  It is important in cavities with nonlinear loss such as silicon-core resonators, and is given by (see Appendix~\ref{FCA:derivation})
\begin{align}
r_\mathrm{FC} &= \frac{\tau_\mathrm{FC}\sigma_a v_g}{2\hbar V_\mathrm{eff}}\left ( \beta_\mathrm{tpa,ss} |A_s|^4 + \beta_\mathrm{tpa,pp} |A_p|^4\right .\nonumber\\
& + \beta_\mathrm{tpa,ii} |A_i|^4 + 4\beta_\mathrm{tpa,sp} |A_s|^2|A_p|^2\nonumber\\
& \left . + 4\beta_\mathrm{tpa,ip}|A_i|^2|A_p|^2 + 4\beta_\mathrm{tpa,si}|A_s|^2|A_i|^2 \right ) \label{FCArate}
\end{align}
where $\tau_\mathrm{FC}$ is the free carrier lifetime, $\sigma_a$ is the free carrier absorption cross section area per electron-hole pair, and $v_g$ is group velocity\footnote{$\sigma_a$ and $v_g$ both only have a meaning in the context of resonators formed from a waveguide, such as microring or waveguide Fabry-Perot resonators.  The expression is still valid for 3D standing wave cavities such as photonic crystal microcavities, where only the product $\sigma_a v_g$ as a whole has a unique physical meaning.}. $V_\mathrm{eff}$ is an effective volume of the resonant mode, as defined in Appendix~\ref{sec:app:ovlintegrals}.

{\bf Approximations and assumptions:} Without loss of generality, we make a few simplifying approximations and assumptions, as follows. Note that, rigorously, there is only one $S_-$ (output) and one $S_+$ (input) port in the system in Fig.~\ref{fig1_schematics}(a), and the above $S_{k,\pm}$ are respective parts of the spectrum of $S_\pm$.  We are making the approximation, relevant to OPO analysis, that the wavelength spacing of the pump, signal and idler resonances is larger than their linewidth, and that we have continuous-wave (CW) operation or nearly so, so that e.g. the signal input wave, $S_{s,+}$, affects only the signal resonance, and does not excite the other two directly, etc.  Then, the three spectral components can be treated as separate ports.  

We further assume that the wavelengths of pump input, $S_{p,+}$, and signal/idler output match the cavity resonances, and that the cavity resonances themselves are spaced to satisfy photon energy conservation in the nonlinear process, $2 \omega_p = \omega_s + \omega_i$.  This is the most efficient case for OPO operation, and is a reasonable assumption because the resonances can either be designed to satisfy frequency matching by dispersion engineering, or can be tuned (e.g. thermally) post-fabrication in some designs. In line with this assumption, we also ignore self- and cross-phase modulation as well as free-carrier induced index change, which result in shifts of the resonance frequencies during operation from their ``cold cavity'' (no excitation) values.  We do this without loss of generality because these shifts can be compensated in principle by pre-shifting the resonance frequencies of the cavity modes in design, or by actively tuning the cavities during operation \cite{Hashemi2009Nonlinear}.

We consider here unseeded operation of an OPO, i.e. free oscillation.  In this case, there is no input power at the signal and idler frequencies beyond noise that is needed to start the FWM process.  We note that,  there are six unique $\beta_\mathrm{tpa,mn}$ coefficients in total (see Eq.~\ref{eqn:loss1}). Therefore a single nonlinear NFOM is not a sufficient metric for performance in integrated photonic structures.  In general, we introduce a $d$-vector to describe the topological mode structure aspects that give rise to differences in the six TPA coefficients.  This is addressed further in Section~\ref{sec:efffom}.  To simplify the analysis and arrive at a single TPA coefficient, $\beta_\mathrm{tpa,mn}\equiv \beta_\mathrm{tpa}$, in some parts of the paper we assume a single, traveling-wave cavity configuration, with traveling wave excitation, and use this for the analysis in Secs.~\ref{sec:pumptpaonly}--\ref{sec:fulltpaandfca}.  In addition, the TPA coefficient $\beta_\mathrm{tpa}$ is defined in Eq.~(\ref{eqn:beta_tpa}) in Appendix~\ref{sec:app:ovlintegrals} and contains the same field overlap integral as $\beta_\mathrm{fwm}$, defined in Eq.~(\ref{eqn:beta_fwm}).  This correspondence allows us to relate our design to the conventional nonlinear figure of merit (NFOM).  Instead of using the NFOM, we find it more natural to work with its inverse, and we define a nonlinear loss parameter, $\sigma_3$, that is a property of the nonlinear material alone for the purposes of this paper.  We call $\sigma_3$ the \emph{nonlinear loss sine} (drawing analogy to the linear loss tangent in electromagnetics) and define it as
\begin{align}
\label{eqn:sigma3def}
\sigma_3 \equiv \frac{\Im[\chi^{(3)}]}{|\chi^{(3)}|}= \frac{\beta_\mathrm{tpa}}{\beta_\mathrm{fwm}}
\end{align}
where we have assumed a scalar $\chi^{(3)}$ (we keep the tensor character in overlap integrals defined later).  $\beta_\mathrm{tpa}$ is referenced to a traveling-wave single-ring design, i.e. uniform field along the cavity length and $\omega_p \approx \omega_s \approx \omega_i$.  
In the general case, $\beta_\mathrm{tpa}$ is replaced by several coefficients $\beta_\mathrm{tpa,pp}$, etc., but $\sigma_3$ can still be defined by the left expression in (\ref{eqn:sigma3def}) via $\chi^{(3)}$.  The nonlinear loss sine $\sigma_3$ depends only on material parameters, and \emph{not} on the overlap integral in the case where a single material in the device dominates nonlinear behavior.  This is because the FWM and TPA have the same overlap dependence.  In the case where multiple nonlinear materials are present in the cavity, the definition of $\sigma_3$ needs to be generalized to include overlap integrals in order to still represent the ratio of two photon absorption to parametric gain.  Since $\sigma_3$ characterizes the relative magnitude of TPA and FWM effects, it is related to the nonlinear figure of merit (NFOM) typically used in the context of nonlinear optical switching,
$\mathrm{NFOM} = \sqrt{1-\sigma_3^2}/(4\pi \sigma_3)$
. For silicon near 1550\,nm, $\sigma_3 \approx 0.23$ \cite{lin2007nonlinear}.

We also assume that each resonance has the same linear loss, $r_\mathrm{k,o} = r_o$.  Last, due to the symmetry of our model in the regime of $\Delta\omega/\omega_p \ll 1$, where $\Delta\omega = \omega_p-\omega_s= \omega_i-\omega_p$ (sufficiently that the signal, idler and pump mode fields confinement is similar), we assume that $\omega_s \approx \omega_i \equiv \omega$, and equal external coupling for the signal and idler resonances, $r_\mathrm{s,ext} = r_\mathrm{i,ext}$.

\section{Normalized model of a parametric oscillator}  \label{sec:normalized_model}

To enable an analysis with more general conclusions, we can rewrite the CMT model in a normalized form:
\begin{subequations}
\begin{align}
\frac{dB_s}{d\tau} & = -\rho_\mathrm{s,tot} B_s - j 2 B_p^2 B_i^{\ast} \label{eqn:dBs}\\
\frac{dB_p}{d\tau} & = -\rho_\mathrm{p,tot} B_p - j 4 B_p^{\ast}B_s B_i - j \sqrt{2\rho_\mathrm{p, ext}}\,T_{p,+} \label{eqn:Bp dynamics}\\
\frac{dB_i}{d\tau} & = -\rho_\mathrm{i,tot}  B_i - j 2 B_p^2 B_s^{\ast}\\
T_{s,-} & = - j \sqrt{2\rho_\mathrm{s, ext}} B_s\\
T_{p,-} & = T_{p,+} - j \sqrt{2\rho_\mathrm{p, ext}} B_p\\
T_{i,-} & = - j \sqrt{2\rho_\mathrm{i, ext}} B_i.
\end{align}
\end{subequations}
The normalized variables are defined by
\begin{subequations}
\begin{align}
\tau &\equiv  r_o t \\
B_k &\equiv \frac{A_k}{A_o} \text{, with } A_o \equiv \sqrt{\frac{2r_o}{\omega\beta_\mathrm{fwm}}} \\
T_{k,\pm} &\equiv \frac{S_{k,\pm}}{S_o} \text{, with } S_o \equiv \sqrt{\frac{2r_o^2}{\omega\beta_\mathrm{fwm}}} \label{eqn:Tpmnormalization}\\
\rho_\mathrm{s,tot} &\equiv 1+ \rho_\mathrm{s,ext} + 2\sigma_3\left(d_\mathrm{ss}|B_s|^2 + 2d_\mathrm{sp}|B_p|^2 + 2d_\mathrm{si}|B_i|^2\right) \nonumber \\
& +\rho_\mathrm{FC} \label{eqn:loss_s_norm}\\
\rho_\mathrm{p,tot} &\equiv 1+ \rho_\mathrm{p,ext} + 2\sigma_3\left(2d_\mathrm{sp}|B_s|^2 + d_\mathrm{pp}|B_p|^2 + 2d_\mathrm{ip}|B_i|^2\right) \nonumber\\
&+\rho_\mathrm{FC} \label{eqn:loss_p_norm}\\
\rho_\mathrm{i,tot} &\equiv 1+ \rho_\mathrm{i,ext} + 2\sigma_3\left(2d_\mathrm{si}|B_s|^2 + 2d_\mathrm{ip}|B_p|^2 + d_\mathrm{ii}|B_i|^2\right) \nonumber\\
&+\rho_\mathrm{FC} \label{eqn:loss_i_norm}.
\end{align}
\end{subequations}
We arrive at normalized energy amplitudes $B_k$ and wave amplitudes $T_{k,+}$, $T_{k,-}$ by normalizing out the linear loss rate $r_o$, parametric coupling $\beta_\mathrm{fwm}$ and nonlinear loss $\beta_\mathrm{tpa}$ from the problem.  Note in Eq.~(\ref{eqn:Tpmnormalization}) that the input/output wave power, $|S_{k,\pm}|^2$, is normalized to $|S_o|^2$ which is the linear-loss oscillation threshold, i.e. the oscillation threshold in the absence of nonlinear losses, as shown later.

The terms $\rho_\mathrm{k,tot} \equiv \frac{r_\mathrm{k,tot}}{r_o}$ and $\rho_\mathrm{k,ext} \equiv \frac{r_\mathrm{k,ext}}{r_o}$, $k \in \{s,p,i\}$ are normalized decay rates.  In order to arrive at an economical formalism to account fully for nonlinear loss, we introduce the nonlinear figure of merit (NFOM), or, more precisely our nonlinear loss sine $\sigma_3$.  In order to preserve the generality required by the six independent $\beta_\mathrm{tpa,mn}$ terms, we introduce the coefficients $d_\mathrm{mn}$, defined as
\begin{align}
d_\mathrm{mn} \equiv \frac{\beta_\mathrm{tpa,mn}}{\sigma_3\beta_\mathrm{fwm}}
\end{align}
which serve as prefactors to the overlap integral ($\beta_\mathrm{tpa}= \sigma_3\beta_\mathrm{fwm}$) of the reference case (i.e. single-cavity with traveling-wave mode).  These six coefficients are a property of the particular resonator topology, and excitation (standing vs. traveling wave), and they together with the NFOM (or $\sigma_3$) completely characterize a device's nonlinear performance merits related to TPA.  In Section~\ref{sec:efffom}, we explain in greater detail these prefactors, calculate them for a few relevant geometries, and discuss the concept of an \emph{effective figure of merit} for four wave mixing in integrated photonic structures.

With the model reduced to a minimum number of coefficients, we last look at the normalized free-carrier absorption rate, given by
\begin{align}
\rho_\mathrm{FC} &\equiv \frac{r_\mathrm{FC}}{r_o} = 	\sigma_3\rho_\mathrm{FC}^\prime \bigl(d_\mathrm{ss} |B_s|^4 + d_\mathrm{pp}|B_p|^4 + d_\mathrm{ii}|B_i|^4 \label{r_norm_full}\\
&+ 4d_\mathrm{sp}|B_s|^2|B_p|^2 +  4d_\mathrm{ip}|B_i|^2|B_p|^2 + 4d_\mathrm{si}|B_s|^2|B_i|^2 \bigr) \nonumber
\end{align}
where we define a normalized FCA coefficient
\begin{align}
\rho_\mathrm{FC}^\prime &\equiv \frac{\tau_\mathrm{FC}\sigma_a v_g}{V_\mathrm{eff}} \frac{\beta_\mathrm{fwm}}{2\hbar}\frac{4r_o}{(\omega\beta_\mathrm{fwm})^2}
= \left(\frac{\sigma_an_\mathrm{nl}^2}{\hbar\omega n_gn_2}\right)\frac{\tau_\mathrm{FC}}{Q_o}.\nonumber
\end{align}
The normalized FCA rate, $\rho_\mathrm{FC}$, depends on nonlinear loss sine $\sigma_3$ (i.e. conventional NFOM), the topological $d$ coefficients, the normalized mode energies ($|B_k|^2$), and a remaining set of parameters lumped into $\rho_\mathrm{FC}^\prime$.  From the last expression, we can see that the FCA effect can be characterized by only one parameter, $\rho_\mathrm{FC}^\prime$, dependent on material nonlinearity, cavity properties and the ratio of free carrier lifetime, $\tau_\mathrm{FC}$, and linear loss Q, $Q_o$.  The last conclusion is interesting even if not entirely surprising -- that free carrier loss depends only on the ratio of free carrier lifetime to the cavity photon lifetime, $\tau_o$, where $Q_o \equiv \omega_o \tau_o/2$.  The larger $\tau_\mathrm{FC}/Q_o$, i.e., $\tau_\mathrm{FC}/\tau_o$, the higher the FCA losses.

The simplifications introduced by this normalized model permit us to numerically solve the optimal synthesis problem for an OPO, which does not have a simple analytical solution, and to arrive at solutions that are universal in the sense that they apply across an array of possible designs.  The model also provides a simple approach to solving similar problems, if certain restrictions we have applied here are removed.  Examples include frequency-mismatched OPO, optical parametric amplifiers and wavelength conversion driven by CW or modulated sources, and parametric spontaneous emission and photon-pair generation.

\section{Finding the optimum OPO design}
\label{sec:this_approach}
Now, we are ready to tackle the ``synthesis'' problem, i.e. the problem of finding the optimum OPO design given certain material parameters.  We define the optimum design of an OPO as the one that, for a given input pump power, provides the maximum output signal (idler) power that can be generated through FWM, i.e. has maximum conversion efficiency.  We define the power conversion efficiency $\eta$ as
\begin{align}
\eta \equiv \frac{|S_\mathrm{s,-}|^2}{|S_\mathrm{p,+}|^2}.
\label{eqn:etadef0}
\end{align}
For similar photon energies, the maximum efficiency is 50\% to each of the signal and idler wavelengths, as two pump photons are converted to one signal and idler photon each (i.e. the maximum photon conversion efficiency is 50\%).

To find the optimum design, the first step is to design resonances for the pump, signal and idler wavelengths that have substantial field overlap and satisfy the energy (frequency) and momentum (propagation constant) conservation conditions (the latter automatically holds for resonances with appropriate choices of resonant orders\cite{kippenberg2004kerr}).  This has been done successfully in previous work \cite{razzari2009cmos, levy2009cmos}.  However, the optimum choice of external coupling and coupled cavity architectures has not been investigated.  We address it here, and show that, in general, unequal waveguide coupling to the pump and signal/idler resonances is the optimum choice, whereas traditional tabletop OPOs typically have equal couplings, as they result from broadband mirrors.
\footnote{Unique exceptions include optical parametric chirped pulse amplification (OPCPA) which explicitly demands mirrors designed for very different coupling at pump and signal wavelengths\cite{ilday2006cavity}.} 

For continuous-wave operation, we begin by finding the steady state conditions of the system.  At steady state ($\frac{d B_k}{d t}=0$) we have
\begin{align}
B_s     &= -2j\rho_\mathrm{s,tot}^{-1} B_p^2 B_i^{\ast}\label{eqn:Bs}\\
B_i     &= -2j\rho_\mathrm{i,tot}^{-1} B_p^2 B_s^{\ast}\label{eqn:Bi}\\
T_{p,+} &= j \frac{ \rho_\mathrm{p,tot} B_p +4j B_p^{\ast}B_s B_i }
{\sqrt{2\rho_\mathrm{p,ext}}}\label{eqn:Tp}
\end{align}
In the following sections, these conditions will lead to the oscillation threshold, efficiency, optimum couplings and other results.  Once the OPO device topology is selected, fixing the topological $d$-vector, the conversion efficiency in Eq.~(\ref{eqn:etadef0}) depends only on 5 normalized parameters, $\eta = \eta(\sigma_3, |T_\mathrm{p,+}|^2, \rho_\mathrm{FC}^\prime, \rho_\mathrm{s,ext}, \rho_\mathrm{p,ext})$.  We proceed to find the design ($\rho_\mathrm{s,ext}, \rho_\mathrm{p,ext}$) with the maximum efficiency $\eta_\mathrm{max}$ for a given normalized input pump power $|T_{p,+}|^2$, nonlinear loss sine $\sigma_3$, and normalized free carrier lifetime $\rho_\mathrm{FC}^\prime$.  The conversion efficiency as a function of external coupling is solved at different levels of simplification below, and optimum couplings are chosen for the maximum efficiency designs.

\section{Traveling-wave single-cavity model with pump-assisted TPA only and no FCA}
\label{sec:pumptpaonly}
As announced in Sec.~\ref{sec:physical_model}, throughout Sections.~\ref{sec:this_approach}--\ref{sec:fulltpaandfca} we assume a single, traveling-wave cavity with a traveling-wave excitation.  In practice, this means three resonant modes with nearly identical time-average spatial intensity patterns (and this is the case in traveling-wave resonators under our assumption of nearby resonance frequencies, which ensure similar transverse waveguide confinement).  In this case, the topological $d$-vector is (see Sec.~\ref{sec:efffom}, Table~\ref{table:cavity_topology})
\begin{align}
d_\mathrm{ss} = d_\mathrm{pp} = d_\mathrm{ii} = d_\mathrm{sp} = d_\mathrm{si} = d_\mathrm{ip} = 1.
\end{align}

First, we solve a simplified version of our model.  We set the nonlinear loss to be dominated by pump-assisted TPA for all three frequencies, and ignore TPA contributions that are much weaker.  That is, we drop the $d_\mathrm{ss}$, $d_\mathrm{ii}$ and $d_\mathrm{si}$ terms from Eqs.~(\ref{eqn:loss_s_norm}) and (\ref{eqn:loss_i_norm}), and drop the $d_\mathrm{sp}$ and $d_\mathrm{ip}$ terms from Eqs.~(\ref{eqn:loss_p_norm}).  This is valid in the weak conversion regime, relevant to many practical situations, where the generated signal and idler light is much weaker than pump light in the cavity. 
In our analysis in this section, we solve this model in all regimes including strong conversion and up to full conversion, even though its physics are valid in the weak conversion regime only.  For one, this analysis provides a useful bound on conversion efficiency.  Then, we also indicate the region of validity for numerical accuracy of the model. In this section we also ignore the loss due to free carrier absorption because it may be effectively reduced by carrier sweep-out using, for example, a reverse biased p-i-n diode \cite{rong2005continuous,dimitropoulos2005lifetime,turner2010ultrashort}.  It is revisited later in the paper.

Thus, the loss rates in Eqs.~(\ref{eqn:loss_s_norm}--\ref{eqn:loss_i_norm}) have the simpler form
\begin{align}
\rho_\mathrm{s,tot} &= 1+ \rho_\mathrm{s,ext} + 4\sigma_3|B_p|^2  \nonumber\\
\rho_\mathrm{p,tot} &= 1+ \rho_\mathrm{p,ext} + 2\sigma_3 |B_p|^2 \nonumber\\
\rho_\mathrm{i,tot} &= 1+ \rho_\mathrm{s,ext} + 4\sigma_3|B_p|^2 
\end{align}
where signal and idler external coupling are equal, as already discussed.  We can find the steady state operating point from Eqs.~(\ref{eqn:Bs}) and (\ref{eqn:Bi}), which gives either
\begin{align}
|B_s|^2 &= |B_i|^2 = 0 & \text{(below threshold)} \nonumber\\
\intertext{or}
\vspace{-6pt}
|B_p|^2 &= \frac{(1+\rho_\mathrm{s,ext})}{2(1-2\sigma_3)} & \text{(above threshold).}\label{eqn:Ap_th}
\end{align}
These two are the classical steady-state solutions to below-threshold and above-threshold operation of the oscillator, respectively.  Note that, at least in the present model that considers pump-assisted TPA only, the steady-state pump resonator-mode energy $|B_p|^2$ is independent of both the input pump power $|T_{p,+}|^2$ and the pump external coupling $\rho_\mathrm{p,ext}$.  This can be interpreted as clamping of the parametric gain, analogous to gain clamping in a laser where, during lasing, the gain saturates to equal the round trip loss (further comments in Appendix~\ref{sec:app:laseranalogy}).  Since the total loss rate (field decay rate) of the freely oscillating mode (signal/idler) includes the external (output) coupling and absorption/radiation losses of the signal/idler resonances only, not those of the pump resonance, it is not surprising that the loss and external coupling of the pump play no role.  The parametric gain is proportional to the cavity pump energy, hence it must stay related to the signal/idler decay rates only, i.e. unchanged when the signal/idler rates are fixed.  In spite of this result, we will show later in this section that the optimum choice of external coupling (both pump and signal/idler) for maximum conversion efficiency does depend on the input pump power, $|T_{p,+}|^2$, i.e. $|S_{p,+}|^2$.

\begin{figure}[htbp]
	\centering	 	
	   \includegraphics[width=3.15 in]{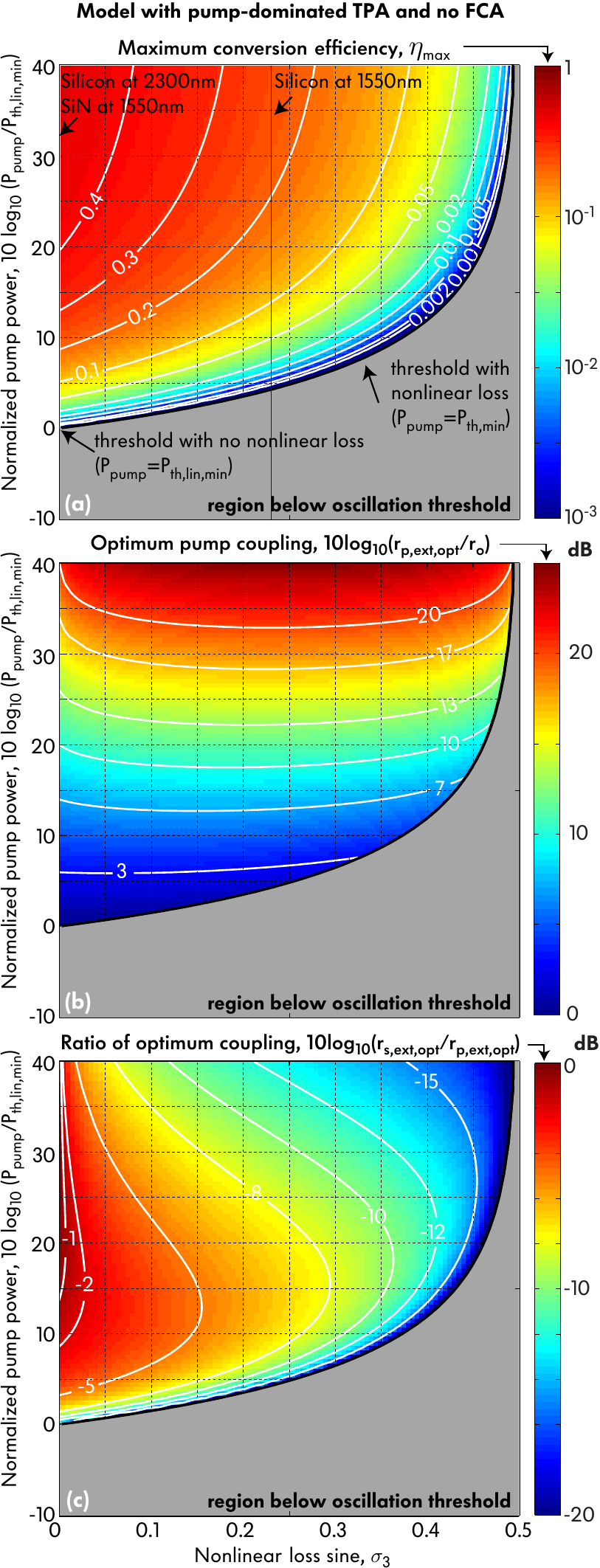}
	   \vspace{-6pt}
	\caption{Normalized design curves for optimum OPO design (model with pump-assisted TPA terms only and no FCA included): (a) maximum efficiency versus pump power and nonlinear loss sine, and corresponding (b) pump resonance coupling and (c) ratio of signal/idler versus pump resonance coupling.\label{fig3_part}}
	\vspace{-3pt}
\end{figure}

We next investigate the oscillation threshold.  In general, the oscillation threshold will depend on the choice of external couplings $\rho_\mathrm{p,ext}$ and $\rho_\mathrm{s,ext}$.  For now, to provide a useful metric for our model normalization, we will study the threshold when we choose the external couplings that \emph{minimize} the oscillation threshold, i.e. give the minimum (optimum) threshold.  This provides a useful and simpler metric that does not depend on couplings.  This minimum threshold pump power, $P_\mathrm{th,min}$, is derived in Appendix~\ref{sec:Threshold pump power}.  It occurs for an external signal coupling of zero, and an external pump coupling set to the nonlinear equivalent of the critical coupling condition.  In the context of the model in this section (without FCA), the minimum threshold is given by
\begin{align}
P_\mathrm{th,min} = \frac{1-\sigma_3}{(1-2\sigma_3)^2}\frac{2r_o^2}{\omega\beta_\mathrm{fwm}} = \frac{1-\sigma_3}{(1-2\sigma_3)^2} P_\mathrm{th,lin,min} \label{eqn:pthpartialtpa}
\end{align}
where $ P_\mathrm{th,lin,min} \equiv 2r_o^2 / (\omega\beta_\mathrm{fwm}) = |S_o|^2$ [compare Eq.~(\ref{eqn:Tpmnormalization})] is the minimum threshold pump power when nonlinear loss is negligible ($\sigma_3 = 0$), and which we'll call the \emph{linear minimum threshold}. The threshold scales as $V_\mathrm{eff}/Q_o^2$ [compare Eq.~(\ref{eqn:pthpartialtpa}) with (\ref{eqn:beta_fwm})], where $Q_o$ is the linear loss Q (unloaded quality factor), and $V_\mathrm{eff}$ is the effective nonlinear mode interaction volume defined in Appendix~\ref{sec:app:ovlintegrals}.  The $\sigma_3$-dependent prefactor in (\ref{eqn:pthpartialtpa}) shows spoiling of the threshold with nonlinear loss, and defines the nonlinear oscillation threshold curve in Figs.~\ref{fig3_part},\ref{fig:allTPA} and \ref{fig5_compare}.  Note that for each NFOM, or nonlinear loss sine $\sigma_3$, which is largely independent of the geometry, the minimum threshold is fixed by a combination of the loss Q and nonlinear interaction volume.  Note also that for $\sigma_3 > 0.5$, the threshold is infinite.  This makes sense -- the two photon absorption is larger than the parametric gain at all pump powers.

Now that the minimum (optimum) threshold is established for all $\sigma_3$, we proceed to find the optimum design at all points above threshold.  Our goal is to express the efficiency only in terms of the input pump power, $\sigma_3$, and the external couplings, which we have control over; then to select the optimum couplings.  This will produce an optimum design for any point in the two-dimensional space of all possible pump powers, and nonlinear loss sine $\sigma_3$ of the material used.

In the steady state, the FWM conversion (pump input to signal output) efficiency $\eta$ in Eq.~(\ref{eqn:etadef0}) can be expressed as
\begin{align}
\eta \equiv \frac{|S_\mathrm{s,-}|^2}{|S_\mathrm{p,+}|^2} = \frac{2r_\mathrm{s, ext} |A_s|^2}{|S_\mathrm{p,+}|^2} =  \frac{2\rho_\mathrm{s,ext} |B_s|^2}{|T_{p,+}|^2}.\label{eqn:etadef}
\end{align}
In this expression, $|B_s|^2$ can be replaced with an expression that depends on $|B_p|^2$ and $|T_\mathrm{p,+}|^2$ using Eqs.~(\ref{eqn:Bi})--(\ref{eqn:Tp}). Then, using Eq.~(\ref{eqn:Ap_th}) we can express $\eta$ as a function only of the input pump power ($|T_\mathrm{p,+}|^2$), external couplings ($\rho_\mathrm{p,ext}$ and $\rho_\mathrm{s,ext}$) and the nonlinear loss sine $\sigma_3$.
 
Next, the maximum efficiency design is found in two steps, first by maximizing efficiency with respect to pump external coupling, and then with respect to signal external coupling.  From $\frac{\partial \eta}{\partial \rho_\mathrm{p,ext}} = 0$ we find optimum solution
\begin{align}
\rho_\mathrm{p,ext,opt} = \frac{1-2\sigma_3}{1+\rho_\mathrm{s,ext}}|T_\mathrm{p,+}|^2.\label{eqn:rpext}
\end{align}
It is straightforward to verify that this choice of pump coupling $\rho_\mathrm{p,ext}$ corresponds to a maximum of $\eta$ for a given input pump power $|T_\mathrm{p,+}|^2$ and signal/idler coupling $\rho_\mathrm{s,ext}$.  We can remove the dependence of $\eta$ on $\rho_\mathrm{p,ext}$ by inserting (\ref{eqn:rpext}) into (\ref{eqn:etadef}). 
Next, setting the derivative of this new $\eta$ with respect to $\rho_\mathrm{s,ext}$ (for a given $|T_\mathrm{p,+}|^2$) to zero, we arrive at a cubic equation in $\rho_\mathrm{s,ext}$:
\begin{align}
(1+\rho_\mathrm{s,ext})^2(2 \sigma_3\rho_\mathrm{s,ext}+1 - \sigma_3) - (1-2 \sigma_3)^2 |T_\mathrm{p,+}|^2 = 0.\nonumber
\end{align}
Since the coefficients of this cubic equation are all real, it always has a real root, given by
\begin{align}
\rho_\mathrm{s,ext,opt} &= \frac{1}{6\sigma_3}(-1-3 \sigma_3+ \left(D - E\right)^{1/3} + \left(D + E\right)^{1/3} ) \nonumber\\ \label{eqn:coupling}
D &\equiv (3\sigma_3-1)^3 + 54 \sigma_3^2(1-2 \sigma_3)^2 |T_\mathrm{p,+}|^2 \nonumber\\
E &\equiv 3 \sigma_3(1-2 \sigma_3) \sqrt{6 |T_\mathrm{p,+}|^2 [(3 \sigma_3-1)^3+D]}
\end{align}
The above solution is only valid when the input pump power, $S_{p,+}|^2$, is above the minimum threshold power $P_\mathrm{th,min}$, which corresponds to the signal coupling $\rho_\mathrm{s,ext,opt}$ in Eq.~(\ref{eqn:coupling}) (and the efficiency $\eta$) taking on positive real values. Again, it can be verified that this solution corresponds to the maximum of $\eta$ in $\rho_\mathrm{s,ext,opt}$ for a given input pump power.  

Thus, in Eqs.~(\ref{eqn:coupling}) and (\ref{eqn:rpext}) we have found a unique optimum design for a parametric oscillator, in closed form, that achieves maximum efficiency $\eta$ for a given ``lossiness'' of the 3$^\mathrm{rd}$-order nonlinearity being used, described by material-dependent nonlinear loss sine $\sigma_3$, and a given input pump power, $|S_{p,+}|^2$.  The design constitutes a particular choice of pump and signal resonance external coupling, providing an optimum conversion efficiency $\eta_\mathrm{max}(|T_{p,+}|^2,\sigma_3) \equiv \eta(|T_{p,+}|^2,\sigma_3,\rho_\mathrm{p,ext,opt},\rho_\mathrm{s,ext,opt})$.  All other parameters that are included in the normalizations ($r_o$, $S_o$ and $A_o$), such as the linear losses, four-wave mixing coefficient, confinement of the optical field, etc., simply scale the solution.

We assume that we can choose couplings freely without affecting other cavity parameters (resonance frequency, loss Q, nonlinear effective volume, etc.) -- thereby separating/decoupling the ``architecture'' (choice of coupled-cavity topology, and coupling between cavities and to waveguide ports) and the building-block ``single cavity design'' (to optimize parametric gain vs. linear radiation loss, sidewall roughness scattering, etc.). Coupling can introduce coupling-strength-dependent parasitic loss\cite{Pop05ofc,Popovic2006mho}, leading to the need to jointly optimize the architecture and cavity design.  These higher-order considerations will be visited in future work.

It is noteworthy that the optimum pump and signal/idler resonance coupling values are different, whereas in practice in table-top parametric oscillators they are typically equal when broadband mirrors are used to set external coupling, e.g. in a Fabry-Perot resonator.

We next study this optimum solution in some detail, starting with a few limiting cases.

\subsection{Optimum designs: lossless $\chi^{(3)}$ ($\sigma_3 = 0$)}
\label{sec:optlosslessc10}
First, we will discuss the limit with no nonlinear loss ($\sigma_3 \to 0$), and then we will examine the full solution just obtained from the pump-assisted-only TPA model.

For $\sigma_3 = 0$, we find that the optimum couplings are
\begin{align}
\rho_\mathrm{s,ext,opt} &= \sqrt{|T_{p,+}|^2} - 1\label{eqn:losslessrsextopt}\\
\rho_\mathrm{p,ext,opt} &= \sqrt{|T_{p,+}|^2}.\label{eqn:losslessrpextopt}
\end{align}
This result is consistent with simple physical intuition.  If the pump power is near the threshold (but above it), i.e. $|T_{p,+}|^2 = 1 + \epsilon$, $\epsilon \ll 1$, then the amount of signal/idler light generated is small, and we are in the undepleted pump scenario.  In this case, the optimum solution is $\rho_\mathrm{p,ext,opt} = 1$, i.e. $r_\mathrm{p,ext,opt} = r_o$, which means that the pump resonance is critically coupled.  Critical coupling maximizes the intra-cavity pump intensity, and hence the parametric gain seen by the signal and idler light.  In the case where the pump power is well above threshold, the generated signal/idler light carries significant energy away from the pump resonance (which acts as a virtual gain medium to the signal/idler light). As a result, the pump resonance sees an additional loss mechanism.  The pump coupling is then larger to match the linear and nonlinear loss combined to achieve ``effective critical coupling'', in which case $\rho_\mathrm{p,ext,opt} > 1$ (i.e. $r_\mathrm{p,ext,opt} > r_o$). For the signal/idler output coupling, near threshold $\rho_\mathrm{s,ext,opt} \approx 0 \ll 1$. Since gain just above threshold exceeds loss by a small amount, the output coupling cannot be large as it would add to the cavity loss and suppress oscillation -- hence, the optimal $r_\mathrm{s,ext}$ is between zero and a small value there.

In the case of far-above-threshold operation, $\sqrt{|T_{p,+}|^2} \gg 1$, and thus $\rho_\mathrm{s,ext,opt} \approx \rho_\mathrm{p,ext,opt} = \sqrt{|T_{p,+}|^2}$.  This also means that $r_\mathrm{s,ext,opt} \approx r_\mathrm{p,ext,opt} \gg r_o$, i.e. the output coupling rate is far above the linear-loss rate.  In the high-power scenario, the optimum design is then equal coupling. This can be understood by analogy to a linear, 2-resonance (second-order) filter\cite{Little97}.  In our analogy, one resonance is the pump and one the signal resonance, coupled by a nonlinearity. If the pumping is strong, and thus the resonance-resonance coupling is large, there is an effective splitting in the modes beyond the intrinsic linewidth $r_o$ (before the gain is included). With well-resolved resonances, for maximum power transfer from the pump input to signal output, symmetric coupling is optimum\cite{dasic2012minimum} (some further comments on analogy to conventional laser in Appendix~\ref{sec:app:laseranalogy}).

In the lossless nonlinearity regime, the optimum design's efficiency (i.e. maximum achievable efficiency) is
\begin{align}
\eta_\mathrm{max}(|T_{p,+}|^2,\sigma_3=0) = \frac{(\sqrt{|T_{p,+}|^2}-1)^2}{2|T_{p,+}|^2}
\end{align}
for $|T_{p,+}|^2 > 1$ (above threshold).  The optimum efficiency together with the corresponding normalized coupling, Eq.~(\ref{eqn:losslessrsextopt})--(\ref{eqn:losslessrpextopt}) provide all of the information needed to design optimum OPOs employing a lossless $\chi^{(3)}$ nonlinearity.

{\bf The forced equal-coupling case ($\rho_\mathrm{p,ext} = \rho_\mathrm{s,ext} \equiv \rho_\mathrm{ext}$):}  For device geometries where different external coupling for different resonances are not easily implemented, we can constrain the pump and signal/idler coupling to all be equal, and can still search for the optimum design in this context.  For each input pump power, $|T_\mathrm{p,+}^\mathrm{(ec)}|^2$, there is an optimum choice of coupling, $\rho_\mathrm{ext}=\rho_\mathrm{ext,opt}$.  Above threshold, this coupling maximizes conversion efficiency.  Near threshold, it equivalently minimizes the threshold power.  The optimum coupling, $\rho_\mathrm{ext,opt}$, is directly related to the pump power by
\begin{align}
|T_\mathrm{p,+}^\mathrm{(ec)}|^2 = \frac{\left(1+\rho_\mathrm{ext,opt}\right)^3\left(1+2\rho_\mathrm{ext,opt}\right)^2}{\rho_\mathrm{ext,opt}\left(3+2\rho_\mathrm{ext,opt}\right)^2} \label{eqn:Tpplusvsrho}.
\end{align}
The normalized oscillation threshold is $\left [ |T_\mathrm{p,+}^\mathrm{(ec)}|^2 \right ]_\mathrm{th} = P_\mathrm{th,min}/P_\mathrm{th,lin,min} = \frac{27}{16}$ (see derivation in Appendix~\ref{sec:Threshold pump power}), and the corresponding normalized external coupling is given by (\ref{eqn:Tpplusvsrho}) as $\rho_\mathrm{ext,opt}=\frac{1}{2}$ at threshold.  This result matches the minimum oscillation threshold expression in Ref.~\onlinecite{kippenberg2004kerr}.  The optimum coupling is exactly half way between the optimum values of $\rho_\mathrm{p,ext,opt}=1$ and $\rho_\mathrm{s,ext,opt}=0$ at threshold in the unconstrained couplings case, described earlier in this section.  At large pump power, $|T_\mathrm{p,+}^\mathrm{(ec)}|^2 \gg 1$, Eq.~(\ref{eqn:Tpplusvsrho}) has an asymptotic form for $\rho_\mathrm{ext,opt} \sim |T_\mathrm{p,+}^\mathrm{(ec)}| - \frac{1}{2}$, which is just the mean value of the optimum couplings in the unconstrained, unequal-couplings case, $\rho_\mathrm{ext,opt} = (\rho_\mathrm{p,ext,opt} + \rho_\mathrm{s,ext,opt})/2$.

\subsection{Optimum designs: with nonlinear loss}
Next, we consider the case with non-zero nonlinear loss, $\sigma_3>0$.  In Fig.~\ref{fig3_part} we plot the maximum efficiency $\eta_\mathrm{max}$, 
and corresponding optimum external coupling rates for the pump and signal/idler, Eqs.~(\ref{eqn:rpext}) and (\ref{eqn:coupling}), as a function of the nonlinear loss sine $\sigma_3$ and normalized input pump power $|T_{p,+}|^2$.  The plots show a few interesting features.  First, the linear losses do not limit the maximum conversion efficiency, but rather merely scale the required pump power and optimum choice of external coupling coefficients.  In the lossless nonlinearity case ($\sigma_3 = 0$), $100\%$ conversion ($\eta=0.5$ to each of the signal and idler) can always be approached with proper design. 
Second, nonlinear loss $\sigma_3$ places an upper limit on the maximum conversion efficiency, increases the threshold, and increases power requirements. Furthermore, oscillation is only possible using nonlinear materials that have $\sigma_3<1/2$. Above this value, the two-photon absorption loss always dominates over the parametric gain, making oscillation impossible.  Even for $\sigma_3 < 1/2$, the two-photon absorption losses set an upper bound on the maximum achievable conversion efficiency, given as (full derivation in Appendix~\ref{Appendix:eta_upper_limit})
\begin{align}
\eta < \frac{1}{2} - \sigma_3.
\end{align}
Note that this is not a tight bound because it results from consideration of pump-assisted TPA only, and an analysis using all TPA contributions will further reduce the maximum conversion and can produce a tighter bound.  Third, a few qualitative characteristics of optimum designs can be seen from the plots.  The optimum pump external coupling is largely independent of the nonlinear loss [see Fig.~\ref{fig3_part}(b)]. On the other hand, the ratio of the optimal signal external coupling to the optimal pump external coupling is largely independent of pump power, and scales primarily with the nonlinear loss [Fig.~\ref{fig3_part}(c)]. 

This model provides useful insight but is numerically accurate for design only for small $\eta$ or for small $\sigma_3$, as discussed later in Sec.~\ref{sec:fulltpanofca} and illustrated in Fig.~\ref{fig5_compare}.  For appreciable output powers (and cavity energies of signal and idler light) or a high nonlinear loss material (large $\sigma_3$), an accurate model of OPO operation requires accounting of all TPA, including that due to resonant signal and idler light.  We consider this more complex model in the next section.

\begin{figure}[t]
    \centering
        \includegraphics[width=3.25 in]{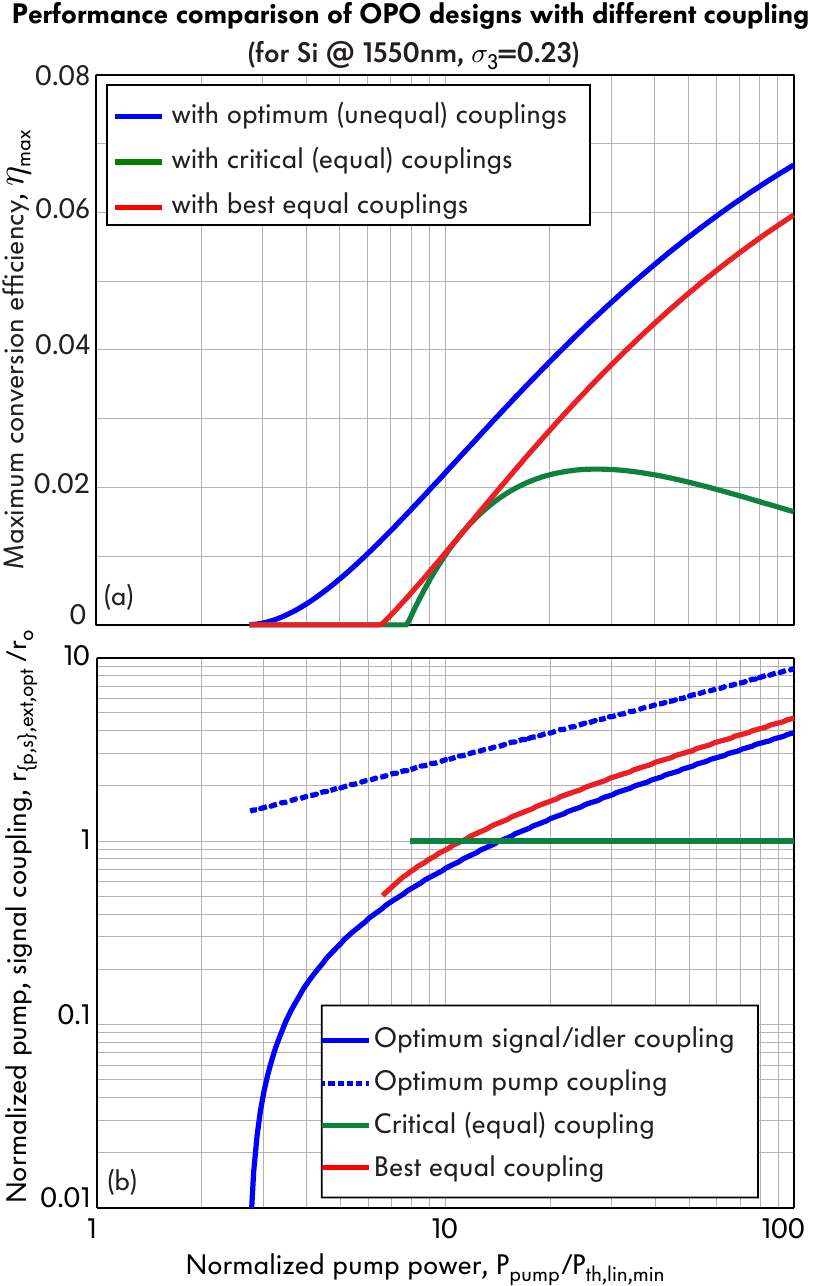}
    \vspace{-6pt}
    \caption{Performance comparison of OPO designs with optimum unequal pump and signal/idler couplings and with optimized equal couplings (assuming no FCA): (a) power conversion efficiency; (b) optimum coupling values.\label{fig2_varCoupling}}
    \vspace{-6pt}
\end{figure}

{\bf The forced equal-coupling case ($\rho_\mathrm{p,ext} = \rho_\mathrm{s,ext}$):}  With nonlinear loss included, the equal-coupling design is again suboptimal.  There is a simple expression for the minimum normalized oscillation threshold power, for the optimum choice of equal couplings, given by (see Appendix \ref{sec:Threshold pump power})
\begin{align}
P_\mathrm{th,min}^{(ec)} = \frac{27(1-\sigma_3)^2}{16(1-2\sigma_3)^3} P_\mathrm{th,lin,min}. \label{eqn:Threshold_ec}
\end{align}
This expression for threshold power is valid in the $\sigma_3 = 0$ case with equal coupling, described in Sec.~\ref{sec:optlosslessc10}.  In Fig.~\ref{fig2_varCoupling}(a), the equal couplings (red) curve crosses the horizontal axis at $6.354$, corresponding to Eq.~(\ref{eqn:Threshold_ec}) with $\sigma_3 = 0.23$ (silicon at 1550\,nm).  The optimum choice of coupling at threshold is still $\rho_\mathrm{p,ext} = \rho_\mathrm{s,ext} \equiv \rho_\mathrm{ext,opt} = \frac{1}{2}$.

To further support that the approach presented here gives the largest conversion efficiency for given input pump power, in Fig.~\ref{fig2_varCoupling} we compare the FWM conversion efficiency of the optimal design to one with all three resonances at the usual critical coupling condition, $\rho_\mathrm{s,ext} = \rho_\mathrm{p,ext} = \rho_\mathrm{i,ext} = 1$.  We also include the case where the couplings are all equal, but are optimized at each value of input power, as calculated above.  The plots show that an unequal coupling design indeed always outperforms one with equal couplings.  Furthermore, it is clear that the critical coupling condition, though it maximizes intracavity pump power and is reasonably close to the optimum design at low powers, is far from optimal for above threshold, and cannot reach maximum conversion efficiency.

\section{Model with full TPA but no FCA}
\label{sec:fulltpanofca}
\begin{figure}[htbp]
	\centering	 	 	 
	   \includegraphics[width=3.15 in]{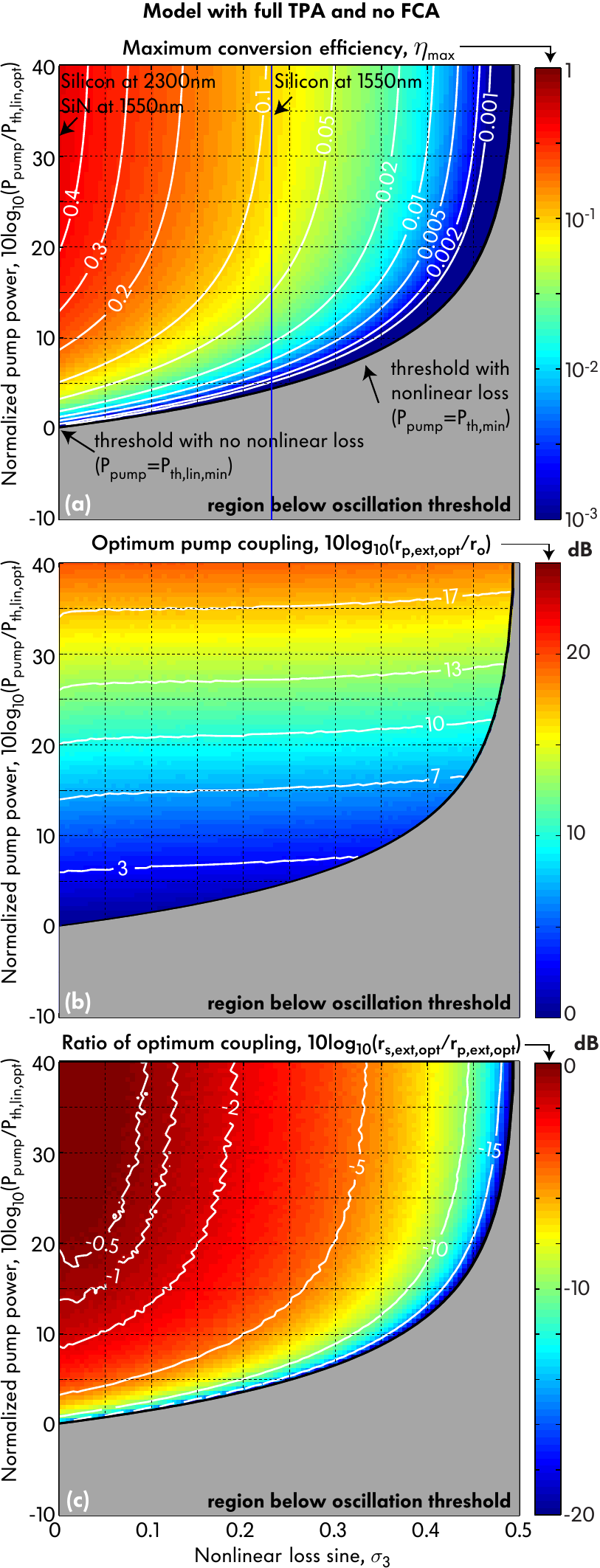}
	   \vspace{-6pt}
	\caption{Normalized design curves for optimum OPO design (model with all TPA terms but no FCA included): (a) maximum efficiency versus pump power and nonlinear loss sine, and corresponding (b) pump resonance coupling and (c) ratio of signal/idler versus pump resonance coupling.\label{fig:allTPA}}
   \vspace{-3pt}
\end{figure}
\begin{figure}[htbp]
	\centering	 	 	 
  	   \includegraphics[width=3.15 in]{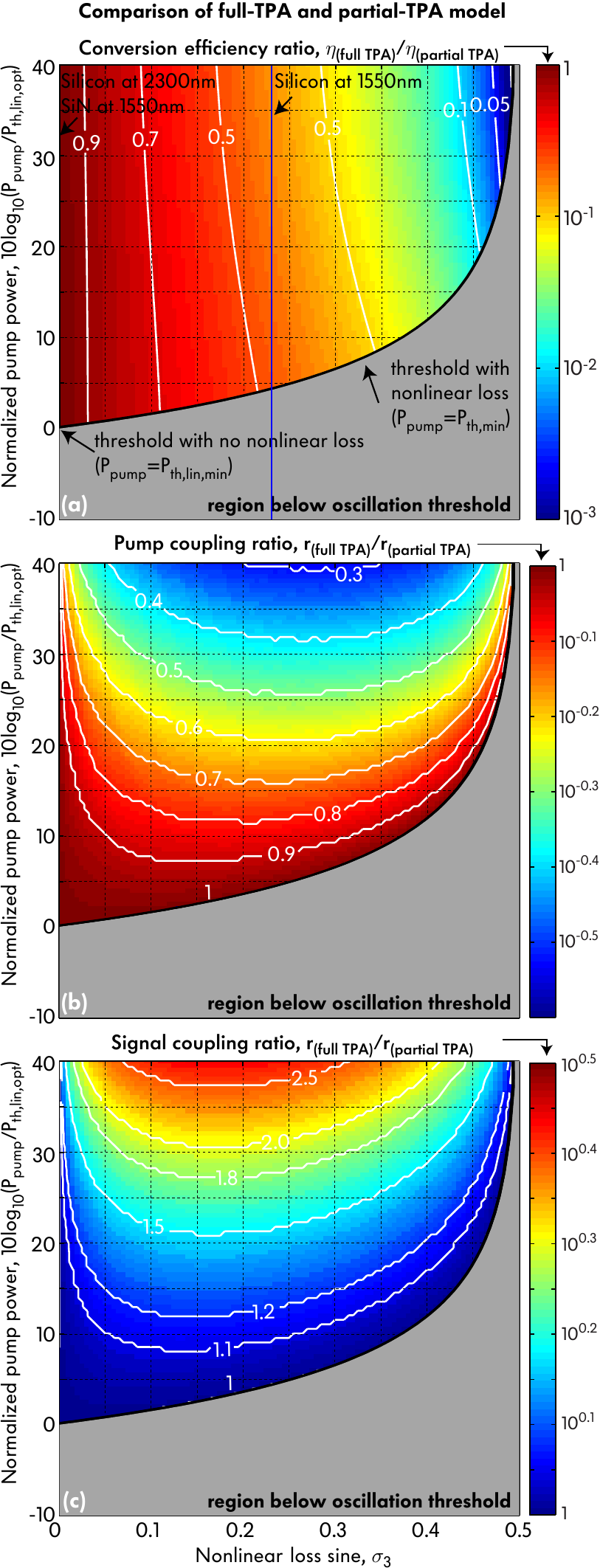}
  	   \vspace{-6pt}
	\caption{Comparison of partial and full TPA model optimum designs (assuming no FCA): (a) maximum efficiency versus pump power and nonlinear loss sine, and corresponding (b) pump and (c) signal/idler resonance external coupling (all values are full TPA case divided by partial TPA case).\label{fig5_compare}}
	\vspace{-3pt}
\end{figure}

In this section we generalize the single-cavity, traveling wave model to include full TPA, including that involving only resonant signal/idler light photons.  This is the complete TPA model needed for systems in the regime of a lossy $\chi^{(3)}$ nonlinearity, with the exception of treatment of FCA, which is deferred to the following section (it is assumed here that the free carrier lifetime can be low enough to not be the limiting loss).  The loss rates in Eqs.~(\ref{eqn:loss1}) have the form
\begin{align}
\rho_\mathrm{s,tot} &= 1+ \rho_\mathrm{s,ext} + 2\sigma_3\left(|B_s|^2 + 2|B_p|^2 + 2|B_i|^2\right)  \nonumber\\
\rho_\mathrm{p,tot} &= 1+ \rho_\mathrm{p,ext}+ 2\sigma_3\left(2|B_s|^2 + |B_p|^2 + 2|B_i|^2\right)  \nonumber\\
\rho_\mathrm{i,tot} &= 1+ \rho_\mathrm{s,ext} + 2\sigma_3\left(2|B_s|^2 + 2|B_p|^2 + |B_i|^2\right)  \label{eqn:loss2)}
\end{align}
There is no longer a simple closed-form expression for the in-cavity steady-state pump light energy as we had in Eq.~(\ref{eqn:Ap_th}) for the partial (pump-assisted-only) TPA model. Instead we have
\begin{align}
|B_p|^2 = \frac{\left(1+\rho_\mathrm{s,ext} + 6\sigma_3|B_s|^2\right)}{2\left(1-2\sigma_3\right)}
\end{align}
which depends on the in-cavity steady state signal/idler light energy, $|B_s|^2$.  It turns out that we need to solve the following cubic equation to find steady-state $|B_s|^2$:
\begin{align}
& 4(1-2 \sigma_3)^3\rho_\mathrm{p,ext}|T_\mathrm{p,+}|^2 = (6\sigma_3|B_s|^2 + 1+ \rho_\mathrm{s,ext})\times \label{eqn:As_fullTPA}\\ 
& \big[(1-2\sigma_3)(1+\rho_\mathrm{p,ext}) + \sigma_3(1+\rho_\mathrm{s,ext})+ 2(2-5\sigma_3^2)|B_s|^2\big]^2\nonumber
\end{align}
and can only then proceed to find the optimum coupling to maximize conversion efficiency.  While it is possible to find an analytical solution for $B_s$ (e.g. with the help of symbolic mathematics software packages\cite{Mathematica}), there is no sufficiently simple analytic expression for it, and we have not succeeded in finding manageable closed form expressions for the optimum couplings themselves in this case.  Nevertheless, we can numerically sweep across values of the parameters $\rho_\mathrm{p,ext}$ and $\rho_\mathrm{s,ext}$ to find the maximum conversion efficiency. This is a worthwhile exercise because it is not computationally expensive, yet the problem is normalized, so a single solution set covers the entire design space. In Fig.~\ref{fig:allTPA}, we show the normalized design curves for optimum OPO design when all TPA terms are included (but no FCA).  We also show a comparison of these two cases -- with partial (pump-assisted-only) TPA and full TPA included, separately -- in Fig.~\ref{fig5_compare}, to show the region of validity of the simpler partial-TPA model.  There is agreement between the partial and full TPA models in efficiency for $\sigma_3 < 0.1$, while the couplings are correct for either pump powers below about $100$ times the minimum nonlinear threshold $P_\mathrm{th,min}$, or for $\sigma_3 < 0.02$ or so, consistent with our comments in the previous section.

Note that the plots in 
Figs.~2--\ref{fig6_Si_Norm} 
imply a \emph{different} optimum device design for each pump power in the sense that the optimum pump and signal/idler coupling are chosen for each value of input power.  In general, for a fixed design, there is an input power which has maximum conversion efficiency, and it is lower at both lower (incomplete conversion) and higher (back-conversion) powers.  Hence, the optimum designs provided are in that sense pump-power specific.


\subsection{Example designs and scaling} \label{sec:ExampleDesigns}
\begin{figure}[t]
	\includegraphics[width=3 in]{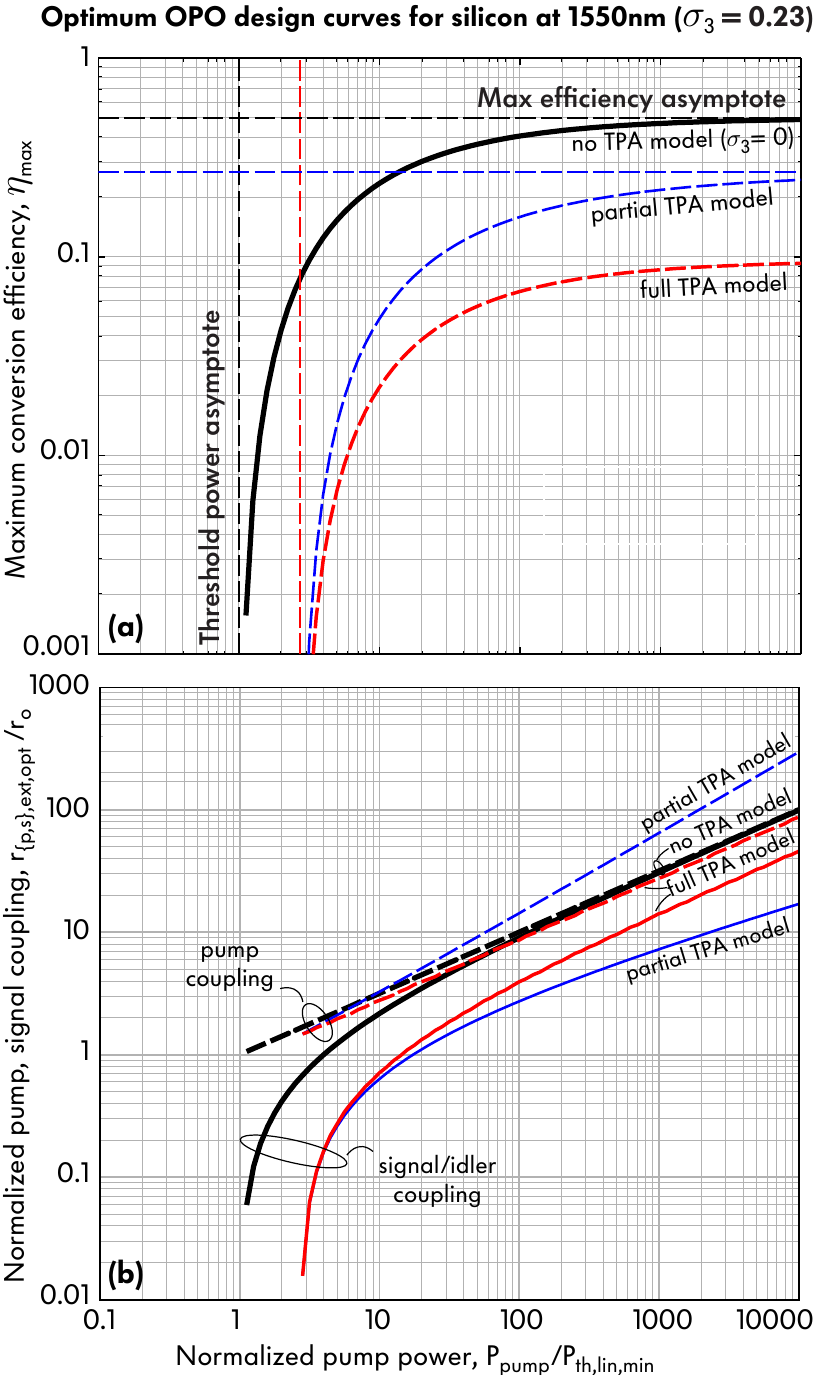}
	\vspace{-6pt}
	\caption{Optimum OPO design curves for nonlinear media with and without TPA loss (assuming no FCA), representative of, e.g., of silicon nitride at 1550\,nm and Si at $2.3\,\mu$m (linear), and Si at 1550 nm ($\sigma_3 = 0.23$).}
	\label{fig6_Si_Norm}
\end{figure}

\begin{table*}[t!]
  \caption{Predicted performance of optical parametric oscillators based on single-ring cavity with traveling-wave  mode}
  \label{table:table1}
	\begin{ruledtabular}
	\begin{tabular}{lccccccccccc} 	 	
Nonlinear & $\lambda$ & $n_2$\footnotemark[2] & $\beta_\mathrm{TPA}$\footnotemark[2] & NFOM & $\sigma_3$ & W$\times$H \footnotemark[3]& $R_\mathrm{out}$ \footnotemark[3]&  $Q_\mathrm{o}$\footnotemark[4] & $V_\mathrm{eff}$ \footnotemark[5]& $\beta_\mathrm{fwm}$ & $P_\mathrm{th}$\footnotemark[7]\\ 
Material\footnotemark[1] &($\mu$m)& $(10^{-5}\frac{\mathrm{cm}^2}{\mathrm{GW}})$ &$(\frac{\mathrm{cm}}{\mathrm{GW}})$ & & & (nm)$\times$(nm) &($\mu$m)& $(10^6)$ & $(\mu$m$^3)$ &  $(10^6$J$^{-1})$ & (mW) \ML  \hline
c-Si &1.55 \cite{lin2007dispersion} & $2.41$ & 0.48  & $0.34$ &0.23   &460$\times$220 &3 & 1 &$2.1$ & $29$ & 0.055 \ML
c-Si &2.3 \cite{lin2007dispersion}& $1.0$ & $\approx 0$ & $\infty$ & $\approx 0$  &700$\times$250 &7 & $1$ &10 &2.5 &0.16 \ML
a-Si:H  &1.55\cite{kuyken2011nonlinear} & 16.6\footnotemark[6] & 0.49\footnotemark[6]& 2.2 & $0.036$  &460$\times$220 & $3$ & $1$ &2.1 &186 &0.004 \ML
Si$_3$N$_4$ &1.55 \cite{levy2009cmos} & $0.24$  & $\approx 0$  & $\infty$  & $\approx 0$ &1600$\times$700 &15 & $1$ &  84 &0.22 & 2.8
	\end{tabular}
	\end{ruledtabular}	
\footnotetext[1]{Defines waveguide core medium only; all devices use silica cladding ($n=1.45$) surrounding the waveguide core.}
\footnotetext[2]{The Kerr coefficient $n_2$ and TPA coefficient $\beta_\mathrm{TPA}$ are related to material $\chi^\mathrm{(3)}$ by\cite{lin2007nonlinear}: $ \frac{\omega}{c}n_2 + \frac{i}{2}\beta_\mathrm{TPA}= 
\frac{3\omega}{4\epsilon_0c^2 n_\mathrm{nl}^2} \chi^\mathrm{(3)}_\mathrm{1111} $.}
\footnotetext[3]{The cavity dimensions in this table are not for actual designs (say, dispersion engineering). We just pick some cavity dimensions to estimate the order of magnitude of threshold power. Here W, H, and $R_\mathrm{out}$ are waveguide core width, height and ring outer radius.}
\footnotetext[4]{$Q_o$ is cavity quality factor due to linear loss only. We assume that $Q_o=10^6$ for all example designs here.}
\footnotetext[5]{$V_\mathrm{eff}$ is effective overlap volume of the signal, pump and idler modes, which are three consecutive longitudinal modes of a microring. Here we ignore waveguide dispersion and assume these three modes have very good overlap in the transverse direction.}
\footnotetext[6]{For a-Si:H, $n_2$ and $\beta_\mathrm{TPA}$ are calculated from Ref.~\onlinecite{kuyken2011nonlinear}. $\beta_\mathrm{TPA}=2A_\mathrm{eff}\gamma_I$, $n2=A_\mathrm{eff}\gamma_R/k_0 $, where $A_\mathrm{eff}$ is mode overlap area, the nonlinear parameter in waveguide $\gamma=770-j28 W^{-1}m^{-1}$.}
\footnotetext[7]{Assuming no FCA; for the case with FCA, see Section~\ref{sec:fulltpaandfca}.}
\end{table*}
We next illustrate use of these design curves.  We use the normalized optimum solution to derive the optimum performance limitations of a few experimentally relevant systems, including OPOs based on silicon and silicon nitride microcavities.  The Kerr (related to parametric gain) and TPA coefficients for crystalline Si and Si$_3$N$_4$ are given in Table~\ref{table:table1}. In the telecom band at 1.55\,$\mu$m wavelength, Si has a large nonlinear loss due to TPA ($\sigma_3 \approx 0.23$ \cite{lin2007nonlinear}), while Si$_3$N$_4$ has negligible TPA ($\sigma_3 \approx 0$) but an order of magnitude smaller Kerr coefficient.  Another promising scenario, pumping silicon above $\lambda \sim 2.2 \mu$m (i.e. photon energy below half the bandgap in silicon), offers both high Kerr coefficient and near zero TPA \cite{lin2007dispersion}.  Also, hydrogenated amorphous silicon \cite{kuyken2011nonlinear} has been shown to have a comparably high NFOM of 2.2 at $\lambda=$1.55\,$\mu$m ($\sigma_3 \approx 0.036$).

In Fig.~\ref{fig6_Si_Norm}, we show slices through Fig.~\ref{fig3_part} and Fig.~\ref{fig:allTPA} showing the still normalized conversion efficiency and corresponding external coupling for optimum designs versus pump power for the 1550\,nm silicon design.  Comparing the partial and full TPA models here again shows agreement at low powers as expected.

To estimate the conversion efficiency and threshold pump power for these reference designs, and to provide some unnormalized example numbers, we assume some typical microring cavity design parameters, given in Table~\ref{table:table1}, and stick to a single ring cavity design.  For example, for a silicon ($n=3.48$)  microring resonant near $\lambda = 1550\,$nm with an outer radius of 3\,$\mu$m, a 460$\times$220 $nm^2$ waveguide core cross-section, surrounded by silica ($n=1.45$), the quality factor of the lowest TE mode due to bending loss is $1.7\!\times\!10^7$.  Considering other linear losses (e.g. sidewall roughness loss), we can assume a total linear loss Q of $10^6$. The effective volume is $2.1\,\mu \mathrm{m}^3$, the FWM coefficient is $\beta_\mathrm{fwm}\approx 2.9\!\times\!10^7\,J^{-1}$, and the minimum linear threshold power, $P_\mathrm{th,lin,min}$, is $21\,\mu$W, while the full minimum nonlinear threshold, with no FCA, is $55\,\mu$W (Table~\ref{table:table1}), and Fig.~\ref{fig7_eta_Si}, discussed later, shows the OPO oscillation threshold when FCA is present.  Operating below half bandgap allows elimination of two-photon losses, but also requires a larger cavity to control linear radiation loss, leading to a similar order threshold in this comparison.  While amorphous silicon\cite{kuyken2011nonlinear}, due to its higher figure of merit, suggests a much lower threshold in this table, in reality achieving linear loss Q's of $10^6$ may not be practical in this material (with measured linear loss of $3.6\,$dB/cm).  And, a:Si-H has been found to degrade over time\cite{kuyken2011nonlinear}.  Due to its weaker nonlinearity and lower index that leads to weaker confinement, silicon nitride suggests thresholds a couple of orders of magnitude higher than silicon.  However, this comparison does not include free carrier losses, discussed in the next section.

\section{Model with full TPA and FCA}
\label{sec:fulltpaandfca}
In a number of $\chi^{(3)}$ materials, including silicon, free carrier absorption (FCA) can be a substantial contributor to optical nonlinear losses.  Hence our results in Sec.~\ref{sec:fulltpanofca} that analyze nonlinear (TPA) loss are numerically valid only when FCA can be neglected, which can occur with sufficient carrier sweepout though this requires strong applied electric fields. In general, with no or incomplete carrier sweepout FCA is present, and must be accounted for.  In this section, we consider solutions to our complete model, including FCA.

To solve for the steady-state in-cavity signal light energy and therefore calculate and maximize conversion efficiency, one needs to first solve for $B_s$ a system of two coupled equations, cubic in $|B_s|^2$ and $|B_p|^2$, respectively.  These are derived from the model in Eqs.~(\ref{eqn:dBs})--(\ref{r_norm_full}).  The steady-state solution for $B_s$ satisfies
\vspace{-6pt}
\begin{align}
\vspace{-12pt}
&\sigma_3\rho_\mathrm{FC}^\prime B_p^4 + \left(8\sigma_3\rho_\mathrm{FC}^\prime B_s^2+4\sigma_3-2\right)B_p^2 \nonumber\\
=& -\left(6\sigma_3\rho_\mathrm{FC}^\prime B_s^4+6\sigma_3B_s^2+1+\rho_\mathrm{s,ext}\right), \\
&\big[(2-2\sigma_3)B_p^2+(4+2\sigma_3)B_s^2+\rho_\mathrm{p,ext}-\rho_\mathrm{s,ext}\big]^2B_p^2 \nonumber\\
=& 2\rho_\mathrm{p,ext,opt}T_\mathrm{p,+}^2
\end{align}
\begin{figure}[t]
	\includegraphics[width=3.25 in]{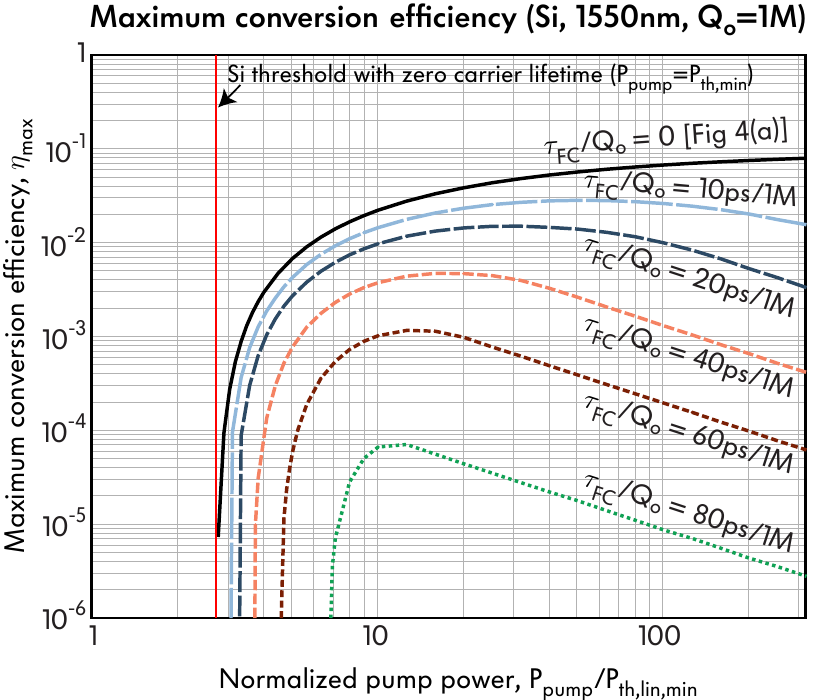}
	\vspace{-6pt}
	\caption{Performance of silicon microcavity at 1550 nm resonance with various free-carrier lifetime and intrinsic cavity quality factors.}
	\label{fig7_eta_Si}
\end{figure}
\normalsize
We have found no simple closed-form analytical expression for $B_s$. In this case, still, one can numerically solve for each $B_s$ and find the optimum coupling for maximum conversion efficiency by sweeping the parameter space.  Because the problem is normalized this is still useful to do, computationally inexpensive and provides a great deal of information.  As can be noted from Eq.~(\ref{r_norm_full}), the loss due to free carrier absorption (FCA), which affects the efficiency $\eta$, scales only with the ratio $\frac{\tau_\mathrm{FC}}{Q_o}$, i.e. the ratio of free carrier lifetime to cavity photon lifetime.  This again simplifies exploration of the design space.

In Fig.~\ref{fig7_eta_Si}, we show simulation results for the silicon microcavity at 1550~nm in Table~\ref{table:table1} with an example set of free carrier lifetimes and cavity loss Q values.  The plot shows that, with FCA present, the optimum design's conversion efficiency, $\eta_\mathrm{max}$, does not monotonically increase with input pump power, as was the case with all of the models in the previous sections (a simple explanation is provided in Appendix~\ref{sec:understandOscThresh}). This is because a stronger pump produces a larger steady-state carrier concentration generated by TPA and, as a result, the overall FCA and total cavity loss is higher at higher pump power.  In fact, the free carrier loss increases faster (quadratically with pump power) than the parametric gain (linearly with pump power), leading to falling conversion efficiency with increasing pump power.  Fig.~\ref{fig7_eta_Si} also shows, however, that even silicon OPOs at 1550\,nm, where TPA and FCA work against the nonlinear conversion process, can achieve conversion efficiencies of ~$0.1\%$ with a pump power of $0.21$\,mW and free carrier lifetime of 60\,ps, which is well within the achievable using carrier sweepout via e.g. a reverse biased p-i-n diode integrated in the optical microcavity\cite{turner2010ultrashort}.  These results and model provide some guidance for future work on efficient implementations of silicon OPOs at 1550\,nm, where TPA and FCA are important.

\begin{figure}[t]
	\begin{center}
	\includegraphics[width=3.25 in]{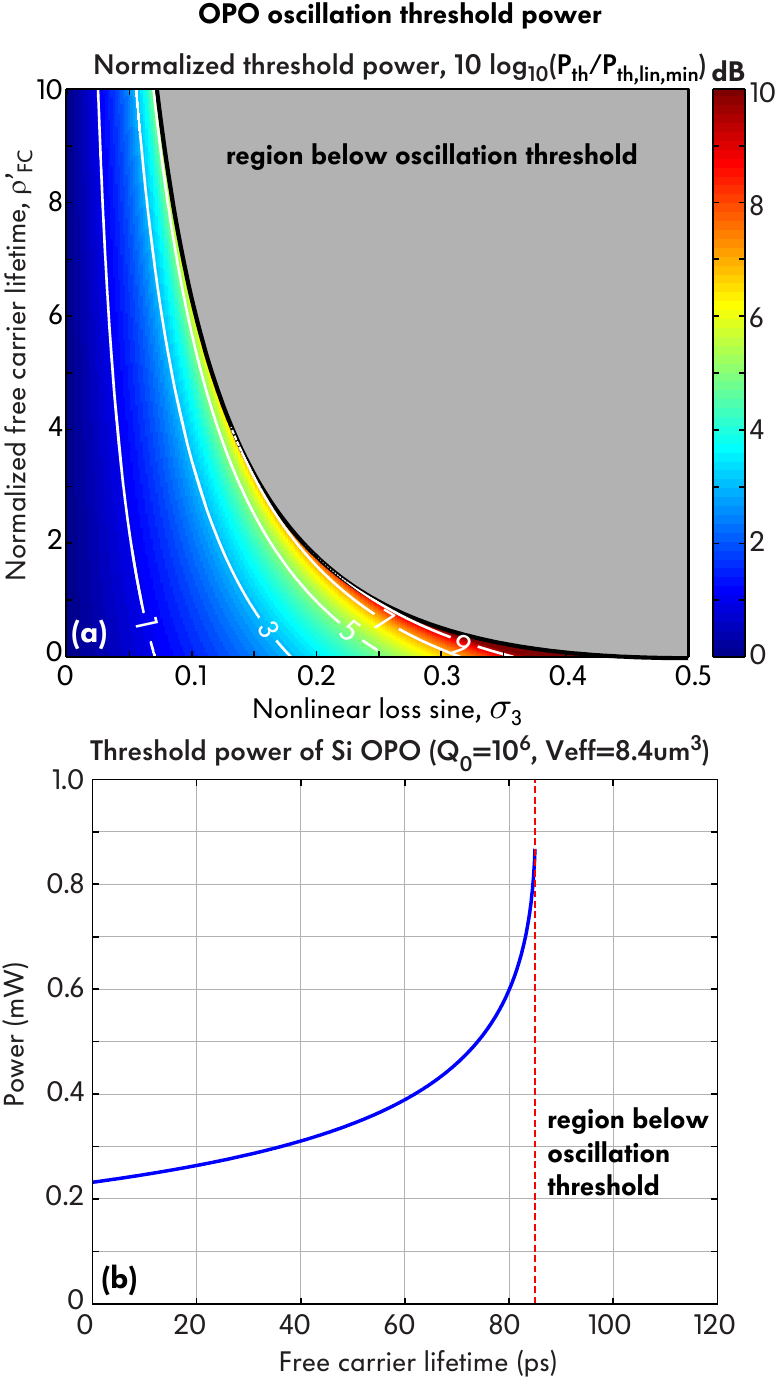}
	\end{center}
	\vspace{-9pt}
	\caption{The OPO threshold vs (a) normalized free carrier lifetime and $\sigma_3$; (b) free carrier lifetime for silicon cavity resonant near 1550 nm with linear unloaded Q of $10^6$ and effective volume of $8.4\,\mu m^3$.}
	\label{fig8_Pth}
	\vspace{-6pt}
\end{figure}

Finally, although the optimum design is not provided in closed form for the model that includes full TPA and FCA, we can derive a closed form expression for the minimum oscillation threshold, which is now different, while it was the same in all previous sections. When FCA is present, the minimum oscillation threshold is (see Appendix~\ref{sec:Threshold pump power}) 
\vspace{-6pt}
\begin{align}
\vspace{-12pt}
P_\mathrm{th,min} = \frac{4(1-\sigma_3)}{\left [ (1-2\sigma_3)+\sqrt{(1-2\sigma_3)^2-\sigma_3 \rho_\mathrm{FC}^\prime}\right ] ^2}P_\mathrm{th,lin,min}
\end{align}
and depends on only the nonlinear loss sine $\sigma_3$, and normalized free-carrier-lifetime, $\rho_\mathrm{FC}^\prime$ (see Eq.~\ref{r_norm_full}). The threshold given by this equation is consistent with the simulation results in Fig.~\ref{fig7_eta_Si}.  In Fig.~\ref{fig8_Pth}(a), we plot the minimum OPO threshold versus the nonlinear loss sine $\sigma_3$ and normalized free-carrier lifetime $\rho_\mathrm{FC}^\prime$. We also plot the minimum OPO threshold for a silicon cavity near 1550\,nm versus actual free carrier lifetime in Fig.~\ref{fig8_Pth}(b), showing that there is a free-carrier lifetime above which oscillation is not possible at any pump power.  We expect this general result for oscillation threshold to also be a useful tool for efficient design of integrated photonic OPOs.

\section{Cavity mode topology and effective figure of merit}
\label{sec:efffom}
In the introduction and coupled-mode theory model (Sections~\ref{sec:introduction}--\ref{sec:normalized_model}), we allowed for general mode and excitation configurations, through the topological $d$-vector.  However, in our analysis thus far, in Sections~\ref{sec:this_approach}--\ref{sec:fulltpaandfca}, for simplicity we chose a single-cavity traveling-wave resonant mode system, e.g. a microring resonator excited in traveling-wave mode.  That analysis is strictly valid if dispersion is such that conversion to resonances adjacent to the pump resonance dominates, i.e. if the system is effectively a three resonance system, but the primary purpose is to illustrate the major parameters that influence the efficiency, optimal choices in design, and performance scaling.

In this section, we address more general resonator configurations, such as that proposed in Fig.~\ref{fig1_schematics}(c), and the degrees of freedom made available through engineering of coupled-cavity compound resonators and of wavelength or mode selective coupling.  This is done by considering the cavity-mode topology $d$-vector, which we introduced in Sec.~\ref{sec:normalized_model}.

First, a triple-cavity resonator such as that in Fig.~\ref{fig1_schematics}(c) is one example of a resonator that explicitly provides only 3 resonant modes near each longitudinal resonance of the constituent microring cavities.  The wavelength spacing of these resonances is determined by ring-ring coupling strength, via the coupling gap\cite{atabaki2011ultra,zeng2013optimum}.  If the dispersion in the building-block microring cavity is sufficiently large, adjacent longitudinal resonances that are spaced 1 free spectral range (FSR) from the utilized resonance will not have proper frequency matching and will not exhibit substantial FWM as a result.  Thus, the nonlinear optics can be confined to the ``local'' three resonances formed by ring coupling at one longitudinal order.  One benefit of the triple-cavity design is that, even if the microring cavity is dispersive and has non-constant FSR, the coupling-induced frequency splitting can be designed to provide equally spaced resonances to enable FWM.  As a result, the individual microring cavity can be optimized for parametric gain, without a competing requirement to produce zero dispersion, while the coupling provides the choice of output signal/idler wavelengths.  By contrast, in a single-ring microcavity, the choice of wavelengths is directly coupled to the size, as is parametric gain, so minimizing the mode volume also requires one to use signal/idler wavelengths that are spaced far apart due to the large FSR, and may put a limit on how small the cavity can be made and still provide a benefit, as dispersion may begin to work against the increase in parametric gain.

\begin{figure}[t]
	\centering	 	 	 
	   \includegraphics[width=3.25 in]{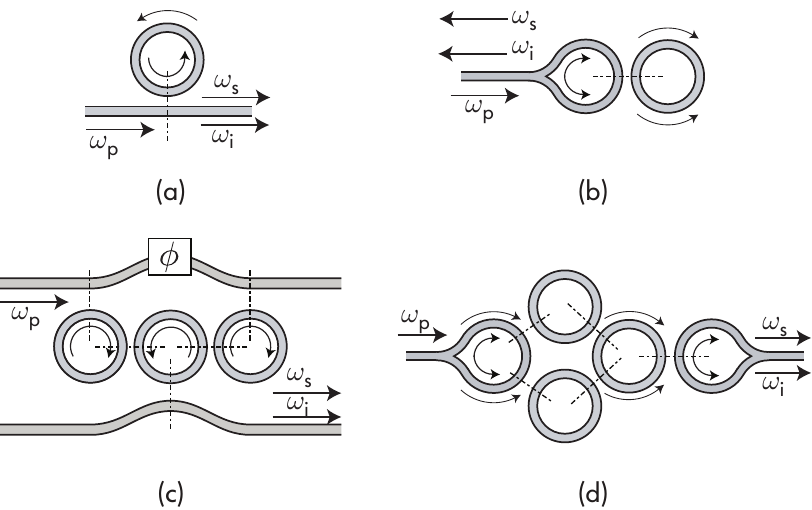}
	\vspace{-6pt}   
	\caption{Effective figure of merit due to mode overlapping in microrings with various structure topology: (a) single-ring cavity with traveling-wave mode; (b)single-ring cavity with standing-wave mode; (c) triple-ring cavity with traveling-wave mode; (d)triple-ring cavity with standing-wave mode.}
	\label{fig10_twswfigA}
	\vspace{-6pt}
\end{figure}
\begin{figure}[t!]
	\centering	 	 	 
	   \includegraphics[width=3.25 in]{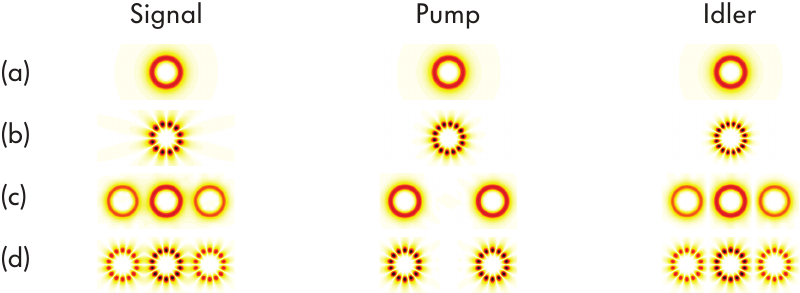}
	\vspace{-6pt}   
	\caption{Mode fields of the pump, signal and idler resonances for the configurations (a)--(d) in Fig.~\ref{fig10_twswfigA} (color coded intensity scale is different in single and triple-cavity cases in order to show the mode features clearly).}
	\label{fig10_twswfigB}
	\vspace{-6pt}
\end{figure}
\begin{figure}[t!]
	\centering	 	 	 
	   \includegraphics[width=3.25 in]{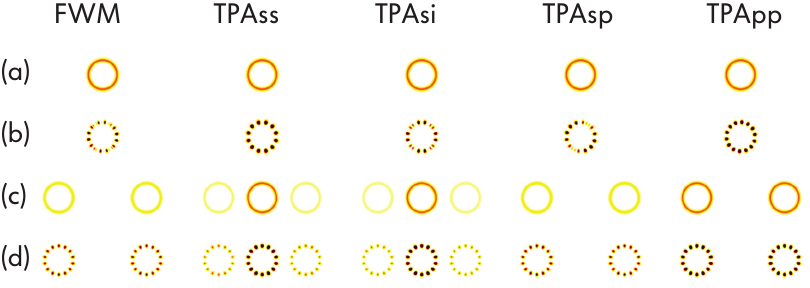}
	\vspace{-6pt}   
	\caption{Mode overlap integrand for the parametric gain and various TPA coefficients for configurations (a)--(d) in Fig.~\ref{fig10_twswfigA}.}
	\label{fig10_twswfigC}
	\vspace{-6pt}
\end{figure}

Compound resonator designs can also provide substantial freedom in engineering of mode selective coupling to the bus waveguide(s).  A single-ring cavity can have wavelength selective coupling, e.g. using a Mach-Zehnder, two-point coupler\cite{popovic2007transparent}.  However, mode interferometry and orthogonal excitation can provide substantial control of coupling in compound resonators, such as that in Fig.~\ref{fig1_schematics}(c).  In the illustration, for example, the bottom waveguide couples only to the signal and idler resonances, because only the signal and idler resonances have non-zero intensity in the middle ring cavity [see Fig.~\ref{fig10_twswfigB}(c)]. 
If the phase shift $\phi$ is chosen so that excitation of the first and last ring are out of phase for the signal/idler resonances, then they will not be excited.  Since the pump resonance has antisymmetric amplitudes in the outer rings, unlike the signal and idler which have symmetric amplitudes, the pump will be efficiently excited by the same configuration, with coupling strength controlled by the choice of coupling gaps.
In this way, the coupling to the pump, and to the signal and idler wavelengths, is entirely decoupled. The waveguide-resonator coupling gap in each case determines the corresponding linewidth, allowing different pump and signal/idler couplings to be implemented, as is required (as shown in earlier sections) to achieve an optimal design.  Note that in this design, in the ideal scenario, the pump resonance is coupled to only the top waveguide, and the signal/idler resonances only to the bottom waveguide.  Therefore, in the linear regime, the top and bottom waveguides are entirely uncoupled, and the only energy coupling from the top to the bottom waveguide can come from nonlinear interaction.  One useful feature is that this design automatically filters the pump.  In practice, with fabrication variations and loss, we would expect 10-20\,dB rejection of the pump to be readily achievable, instead of complete decoupling, but this can still provide a useful function in parametric oscillators.

\begin{table}[t]
  \caption{Comparison of FWM and TPA coefficients in various cavity topologies}
  \label{table:cavity_topology}
	\begin{ruledtabular}
		\begin{tabular}{ l || c| c|c|c|c} 
			Cavity Type\footnotemark[1] & $\frac{\beta_\mathrm{fwm}}{\beta_\mathrm{fwm}\text{(1-ring,TW)}}$\footnotemark[2] & $d_\mathrm{ss}$ & $d_\mathrm{pp}$ & $d_\mathrm{sp}$ &  $d_\mathrm{si}$ \\ 
		    \hline
		    1-ring (TW\footnotemark[3])  & $1$ 				& $1$	      & $1$           & $1$               & $1$			\\
		    3-ring (TW\footnotemark[3])	 & $\frac{1}{4}$	& $\frac{3}{2}$  	  & $2$  	  & $1$  	       & $\frac{3}{2}$ \\    
		    1-ring (SW\footnotemark[3])  & $\frac{1}{2}$	& $3$	      & $3$           & $2$               & $2$			\\
		    3-ring (SW\footnotemark[3])	 & $\frac{3}{8}$    & $\frac{3}{2}$  	  & $2$  	  & $1$  	       & $\frac{3}{2}$ 	\\    
   		\end{tabular}
	\end{ruledtabular}
	\footnotetext[1]{Each constituent ring of the triple-ring cavity is identical to the single-ring cavity.}
	\footnotetext[2] {Four wave mixing coefficients are normalized to that of a single-ring cavity with traveling-wave modes.}
	\footnotetext[3]{TW: traveling-wave, SW: standing-wave.}
\end{table}

Next, we show how these geometries can be simply included in our analysis formalism, through the $d$-vector and an effective figure of merit.  This allows a generalized approach to design, that produces a normalized optimal solution for each unique geometry subspace with a given $d$-vector and NFOM.

To simplify the analysis, we replace the six distinct TPA coefficients, $\beta_\mathrm{tpa,mn}$, by four -- after making the assumption that the frequency splitting $\Delta\omega = \omega_p-\omega_s= \omega_i-\omega_p$ is small, and that signal and idler mode confinement is similar, hence the effective two-photon absorption coefficients are about the same ($\beta_\mathrm{tpa,ii} = \beta_\mathrm{tpa,ss}$ and $\beta_\mathrm{tpa,ip} = \beta_\mathrm{tpa,sp}$).  

As shown in Appendix~\ref{sec:app:ovlintegrals}, the four wave mixing coefficients $\beta_\mathrm{fwm,k}$ ($k \in \{s,p,i\}$) and two photon absorption coefficients $\beta_\mathrm{tpa,mn}$ ($m,n \in \{s,p,i\}$), are dependent on the overlap integral of interacting cavity modes. For different cavities, these coefficients and their relative magnitudes can be very different. As an example, in Fig.~\ref{fig10_twswfigA} we show resonators consisting of a single ring cavity or triple coupled ring cavities, each with either traveling-wave mode or standing-wave mode excitation.  The supermodes are shown in Fig.~\ref{fig10_twswfigB}.  In Fig.~\ref{fig10_twswfigC}, we plot the parametric gain due to FWM and loss due to TPA in these microcavities.
It turns out that, for example, the ratio of signal-idler TPA to parametric gain (i.e. $\sigma_3d_\mathrm{si}$) is larger in a triple-ring resonator than in a single-ring resonator, with a traveling-wave mode excitation.  This means that the effective figure of merit of the triple-ring resonator is smaller than that of the single-ring cavity.  A complete summary of various FWM coefficients and $d$-vectors is shown in Table~\ref{table:cavity_topology}.

Table~\ref{table:table3ring} shows the results of Table~\ref{table:table1} evaluated for a 3-coupled cavity ``photonic molecule'' OPO with traveling-wave excitation, based on the same ring cavity design in each case.

The important conclusion from this study is that the cavity \emph{envelope} matters, i.e. the distribution of the field across parts of the compound resonator, as well as standing vs. traveling wave excitation.  Specifically, standing-wave excitation is very ``efficient'' for self-TPA loss terms, such as absorption of two signal photons or two pump photons.  On the other hand, because of differences in longitudinal mode order, the parametric gain is a bit suppressed.  Thus, standing wave excitation in general loses to traveling-wave excitation in the presence of TPA.  Likewise, the single-ring configuration is more efficient than triple ring with traveling-wave excitation. However, with standing-wave excitation, the single-ring resonator has a larger FWM coefficient but at the same time larger TPA loss ($d$ coefficients). However, it should be kept in mind that this comparison is for microring cavities being equal.  The triple ring design may be able to use much smaller ring cavities than a single ring design, however, as it is not limited by dispersion.  Therefore, either may be more efficient, depending on specific implementation, target wavelengths, etc.

\begin{table}[t]
  \caption{Predicted performance of optical parametric oscillators based on 3-ring photonic molecule with traveling-wave mode } 
  \label{table:table3ring}
	\begin{ruledtabular}
	\begin{tabular}{lcccccccccc} 	 	
Nonlinear & $\lambda$ & $V_\mathrm{eff}$ \footnotemark[1]& $\beta_\mathrm{fwm}$ & $P_\mathrm{th}$\\ 
Material &($\mu$m)& $(\mu$m$^3)$ &  $(10^6$J$^{-1})$ & (mW) \ML  \hline
c-Si &1.55& 8.3 & 7.2 & 0.29 \ML
c-Si &2.3& 40 & 0.63 & 0.65 \ML
s-Si:H &1.55 & 8.3 & 46 & 0.015 \ML
Si$_3$N$_4$ &1.55& 337 & 0.05 & 11
	\end{tabular}
	\end{ruledtabular}	
\footnotetext[1] {Each constituent ring of the triple-ring cavity is identical to the single-ring cavity in Table~\ref{table:table1}.}
\end{table}

More generally, our conclusions are that higher-order resonator designs can provide both unique functionality, and access to degrees of freedom needed to produce an optimum design.  The model presented provides a normalized solution vs. normalized pump power (that includes linear losses), nonlinear FOM $\sigma_3$ and a normalized FCA, for each resonator ``topology'' with a unique $d$-vector.  This should provide a basis for exploring efficient device designs and novel applications.

\section{Future work}

The purpose of this paper was to lay the theoretical foundation for designing efficient parametric oscillators, and for considering the degrees of freedom made available in design by complex photonic structures such as coupled-cavity resonators.  A number of details will play an important role in determining the practical utility of these designs.  For example, we have here assumed that ring-ring coupling is lossless.  In practice, couplers exhibit radiation loss\cite{popovic2005high,popovic2006multistage} and will limit the performance of coupled-cavity designs for FWM.  Design details such as this will be specific to particular implementations, and are left for future study.

Furthermore, we considered the optimal case here with perfect frequency matching.  In an experimental situation, a frequency mismatch must be admitted, and this requires only a simple modification of the presented model.

More generally, the degrees of freedom available in coupled resonant structures on chip suggest that complex synthesis and designs will enable either optimal designs or ones with unique capability for other applications, such as parametric amplifiers and entangled photon sources, including design of joint spectral and temporal distribution of the bi-photons, their coincidence properties, etc.

The distributed states formed in coupled-cavity resonators suggest an analogy to distributed electronic states in molecules, so we sometimes refer to these structures as ``photonic molecules''.  What is unique, in comparison to atomic molecules, is that here we can engineer the resonant frequencies (energies) of the distributed states, their overlap integrals (for purposes of nonlinear optics) by choice of coupling geometries, their effective ``absorption and emission cross-sections'' to waveguides via choice of coupling.  Beyond degrees of freedom available in natural molecules, we can further design anisotropic absorption and emission, i.e. couple different wavelength resonances to entirely separate radiation channels of waveguides.

More generally, there exist coupled cavity geometries that provide unique state manipulation that is either non-existent or unusual in natural atomic molecules, such as radiative coupling which leads to energy-level attraction, in contrast to the usual energy level repulsion\cite{dahlem2010dynamical,gentry2013dark}.  Such devices can even have splitting in the radiative lifetime instead of energy level, i.e. resonant frequency.  Such designs likewise may have applications in single photon sources and unique types of laser cavities\cite{gentry2013dark}.

\section{Conclusion}
Generally, the results in this paper show that efficient micro-OPOs can be designed in the presence of only linear losses, and even with limited nonlinear losses, as well as free carrier absorption.  Notably, while devices without TPA call for equal optimum external coupling to all three resonances when the pump is far above threshold, we show that both in the case closer to the threshold, and in the case with substantial nonlinear losses, it is necessary to design substantially different signal/idler, and pump resonance external coupling for optimum performance. 
In the case where nonlinear losses are present, we have shown that a large set of practical cases can be solved by considering only pump-induced TPA.  In this case, we provide an analytical solution to the design.  With full (pump and signal/idler induced) TPA, and with FCA loss included, more complex models do not admit simple closed-form solutions.  However, we  provide a normalized set of design equations, based on which we can numerically solve for the optimum design, and provide a single set of normalized design plots relevant for design using all nonlinear materials, linear cavity losses, and pump powers.

These results have also motivated our proposal of both spatial mode and Q engineered multimode resonators, based on multiple coupled cavities, for nonlinear FWM applications \cite{zeng2013optimum}.  The requirements of an optimum OPO design presented here, primarily the different external coupling for pump and signal/idler resonances suggested by the results, do not fit well with a simple linear cavity such as a Fabry-Perot resonator, or a microring cavity, with broadband coupling to an external excitation port (via a mirror, or directional coupler, respectively).  Yet, such designs have been common in tabletop OPOs \cite{boyd2008nonlinear} as well as on-chip OPOs in silicon and silicon nitride \cite{levy2009cmos,kippenberg2004kerr}.  We believe this analysis suggests further work on more advanced geometries (such as Fig.~\ref{fig1_schematics}(c)) may enable more efficient designs, and may enable one to reach the performance bounds found in this paper.

This work also suggests that complex photonic circuits may provide useful solutions not only for OPOs but also for other devices including parametric amplifiers and entangled photon sources.  A large amount of research was done in the early to mid 1900's in electrical circuit linear filter synthesis using resistors, capacitors and inductors, leading to a body of sophisticated linear filter design techniques.  Development of synthesis of nonlinear circuits based on resonators as building block components may yield a similarly rich array of solutions to nonlinear optics design on chip.

\begin{acknowledgments}
We thank Prof. Kelvin Wagner at University of Colorado Boulder for helpful discussions, and Cale Gentry for input on the manuscript.  This work was supported in part by start-up funds from the College of Engineering and Applied Science, University of Colorado Boulder, and by the David \& Lucile Packard Foundation through a Packard Fellowship in Science \& Engineering.
\end{acknowledgments}

\appendix
\section{Mode overlap integrals} \label{sec:app:ovlintegrals}
For plane wave propagating in bulk medium, the FWM coefficient, $\beta_\mathrm{fwm}$, is directly related to the third-order susceptibility of the nonlinear material, $\overline{\overline{\chi}}^{(3)}$.  In a microphotonic structure, the optical fields are tightly confined and the FWM coefficient also depends on an overlap integral of the interacting mode fields, given by \cite{Rodriguez2007chi2,lin2007nonlinear}
\vspace{-6pt}
\begin{align}
\beta_\mathrm{fwm,s} 
&=  \frac{\frac{3}{16}\epsilon_0\int{d^3{\bf x}\left({\bf E}_s^\ast\cdot{\bf \overline{\overline{\chi}}^{(3)}}:{\bf E}_p^2{\bf E}^\ast_i\right)}}{\sqrt{\int{d^3{\bf x}\left(\frac{1}{2}\epsilon |{\bf E}_s|^2\right)}
\int{d^3{\bf x}\left(\frac{1}{2}\epsilon |{\bf E}_i|^2\right)}}\int{d^3{\bf x}\left(\frac{1}{2}\epsilon |{\bf 	E}_p|^2\right)}} \nonumber \\ 
&\equiv \frac{3\chi^\mathrm{(3)}_\mathrm{1111}}{4n_\mathrm{nl}^4\epsilon_0 V_\mathrm{eff}} \label{eqn:beta_fwm}
\end{align}
where $n_\mathrm{nl}$ is the refractive index of nonlinear material, $\epsilon_0$ is vacuum permittivity, $V_\mathrm{eff}$ is effective volume given by 
\begin{align}
V_\mathrm{eff}\equiv \frac{\chi^\mathrm{(3)}_\mathrm{1111}{\sqrt{\int{d^3{\bf x}\left(\epsilon |{\bf E}_s|^2\right)}\int{d^3{\bf x}\left(\epsilon |{\bf E}_i|^2\right)}}\int{d^3{\bf x}\left(\epsilon |{\bf E}_p|^2\right)}}}{\epsilon_0^2n_{nl}^4\int{d^3{\bf x}\left({\bf E}_s^\ast\cdot{\bf \overline{\overline{\chi}}^{(3)}}:{\bf E}_p^2{\bf E}^\ast_i\right)}}.
\end{align}
The effective volume, $V_\mathrm{eff}$, is the equivalent bulk volume of nonlinear medium, in which uniform fields with the same energy would have equal nonlinearity ($\beta_\mathrm{fwm}$). With the full permutation symmetry of $\overline{\overline{\chi}}^{(3)}$, we have $\beta_\mathrm{fwm,s} = \beta_\mathrm{fwm,i} =\beta_\mathrm{fwm,p}^\ast$ (the Manley-Rowe relations). 

The nonlinear loss coefficients due to two-photon absorption, $\beta_\mathrm{tpa,mn}$ (due to absorption of a photon each from modes $m$ and $n$, $m, n \in \{s, p, i\}$), are described by similar overlap integral. For example,
\begin{align}
\beta_\mathrm{tpa,sp} =  \frac{\frac{3}{16}\epsilon_0\int{d^3{\bf x}\left({\bf E}_s^\ast\cdot\Im[{\bf \overline{\overline{\chi}}^{(3)}}]:{\bf E}_s{\bf E}_p{\bf E}_p^\ast\right)}}{\int{d^3{\bf x}\left(\frac{1}{2}\epsilon |{\bf E}_s|^2\right)}\int{d^3{\bf x}\left(\frac{1}{2}\epsilon |{\bf E}_p|^2\right)}}. \label{eqn:beta_tpa}
\end{align}

\section{Free carrier absorption rate} \label{FCA:derivation}
Here we derive the loss rate of cavity mode amplitude envelop ($A_k$ in the CMT model, for $k \in {s,p,i}$) due to free carrier absorption induced by two-photon absorption (see Eq.~(\ref{FCArate})). On the one hand, free carriers are created through TPA with equal densities. In general, the dynamics of free carrier density, $N_\nu$, is governed by the continuity equation \cite{sze2006physics}
\begin{align}
\frac{\partial N_\nu}{\partial t} &= G- \frac{N_\nu}{\tau_\nu}+ D_\nu \nabla^2N_\nu- s_\nu\mu_\nu{\bf \nabla}\cdot(N_\nu {\bf E}_\mathrm{dc}) \nonumber \\
 &\equiv G- \frac{N_\nu}{\tau_{\nu,\mathrm{eff}}}
\end{align}
where $\nu = e$ for electrons, $\nu = h$ for holes, $s_h=1$, $s_e=−-1$, $D_\nu$ is the diffusion coefficient, $\mu_\nu$ is the mobility, ${\bf E}_\mathrm{dc}$ is applied dc electric field, $\tau_\nu$ is the carrier lifetime, and $\tau_{\nu,\mathrm{eff}}$ is the effective carrier lifetime that includes all the effects of recombination, diffusion and drift. G is the free carrier generation rate per volume due to TPA, where one pair of electron and hole is generated for every two photons absorbed
\begin{align}
G &= \frac{1}{2\hbar\omega}\frac{\Delta E}{\Delta t\cdot \Delta V} = \frac{1}{4\hbar\omega}\Re[{{\bf E}^\ast_\mathrm{tot} \cdot {\bf J}}] \nonumber \\
  &= \frac{1}{4\hbar\omega}\Re[j\omega\epsilon_0 {{\bf E}^\ast_\mathrm{tot} \cdot \overline{\overline{\chi}}^{(3)} : {\bf E}^3_\mathrm{tot}}]
\end{align}
where ${\bf E}_\mathrm{tot}$ is total electric field (${\bf E}_\mathrm{tot}={\bf E}_\mathrm{s}+{\bf E}_\mathrm{p}+{\bf E}_\mathrm{i}$). Thus the steady-state free carrier density is given by
\begin{align}
N_\nu = G \tau_{\nu,\mathrm{eff}}.
\end{align}
On the other hand, these free carriers contribute to optical loss. The free carrier absorption coefficient of optical power (absorption rate per distance) is 
\begin{align}
\alpha_\nu = \sigma_\nu N_\nu 
\end{align}
where $\sigma_\nu$ is free carrier absorption cross section area. Note that both $\tau_{\nu,\mathrm{eff}}$ and G are position-dependent, and therefore the free carrier absorption coefficient $\alpha_\nu$ is non-uniform across the waveguide cross section. Besides, the optical field intensity is also non-uniform. As a result, the interplay between free carriers and optical field needs to be studied carefully. If the field decay rate due to free carrier loss is much smaller than the cavity resonance frequency, we can include the FCA loss into the perturbation theory of CMT model, with the free carrier loss rate of mode $k$ (for $k \in {s,p,i}$) due to free carrier $\nu$ as
\begin{align}
r_\mathrm{k,FC}^\mathrm{\nu} =& -\frac{j\omega}{4}\frac{\int{d^3{\bf x}\left({\bf E}^\ast_\mathrm{k}\cdot \delta{\bf P}^\mathrm{(FCA,\nu)}_\mathrm{k}\right)}} {\int{d^3{\bf x}\left(\frac{1}{2}\epsilon |{\bf E_\mathrm{k}}|^2\right)}} \nonumber \\
=& \frac{\omega}{4}\frac{\int{d^3{\bf x}\left(\epsilon_0 n_\mathrm{nl} \frac{\alpha_\mathrm{\nu}}{k_0} |{\bf E}_\mathrm{k}|^2 \right)}} {\int{d^3{\bf x}\left(\frac{1}{2}\epsilon |{\bf E}_\mathrm{k}|^2\right)}} \nonumber \\
=& \frac{\epsilon_0 n_\mathrm{nl} \omega \sigma_\nu}{4k_0}\frac{\int{d^3{\bf x}\left( G\tau_\mathrm{\nu,eff} |{\bf E}_\mathrm{k}|^2 \right)}} {\int{d^3{\bf x}\left(\frac{1}{2}\epsilon |{\bf E}_\mathrm{k}|^2\right)}} \nonumber \\
=& \frac{c\epsilon_0^2 n_\mathrm{nl}\sigma_\nu} {16\hbar}
\frac{\int{d^3{\bf x}\left( \tau_\mathrm{\nu,eff}
({\bf E}^\ast_\mathrm{tot} \cdot \Im[\overline{\overline{\chi}}^{(3)}]:{\bf E}^3_\mathrm{tot})
 |{\bf E}_\mathrm{k}|^2 \right)}} {\int{d^3{\bf x}\left(\frac{1}{2}\epsilon |{\bf E}_\mathrm{k}|^2\right)}}. \nonumber \\ \label{r_FC}
\end{align}
This expression for free carrier absorption rate is true, but very complex to solve. We make some assumptions to simplify the expression above. First, we assume the effective free carrier lifetime, $\tau_\mathrm{\nu,eff}$, is the same for electrons and holes. Second, we assume the steady-state free carrier density generated by TPA is uniform (invariant with respect to position) in the cavity.  This assumption is valid when the carrier density equilibrates due to a diffusion that is much faster than recombination\cite{motamedi2012ultrafast}, or a fast drift due to an applied field for carrier sweepout.
With these assumptions, we use the effective volume of nonlinear interaction, $V_\mathrm{eff}$ (defined in Appendix~\ref{sec:app:ovlintegrals}), to average out the free carrier density, $N_\nu$.  From Eq.~(\ref{eqn:dtAs})-(\ref{eqn:dtAi}), using $\frac{\partial N_\nu}{dt} =-\frac{N_\nu}{\tau_\mathrm{eff}}+ \frac{1}{2\hbar\omega V_\mathrm{eff}}\frac{d|A_k|^2}{dt}=0$, 
\begin{align}
N_\nu &= \frac{\tau_{\mathrm{eff}}}{\hbar V_\mathrm{eff}}\left ( \beta_\mathrm{tpa,ss} |A_s|^4 + \beta_\mathrm{tpa,pp} |A_p|^4\right .\nonumber\\
& + \beta_\mathrm{tpa,ii} |A_i|^4 + 4\beta_\mathrm{tpa,sp} |A_s|^2|A_p|^2\nonumber\\
& \left . + 4\beta_\mathrm{tpa,ip}|A_i|^2|A_p|^2 + 4\beta_\mathrm{tpa,si}|A_s|^2|A_i|^2 \right ). \label{FreeCarrierGeneration}
\end{align}
The optical field decay rate due to FCA is given by
\begin{align}
r_\mathrm{FC} =& \frac{\alpha_\mathrm{FC}v_g}{2} = \frac{\sigma_\mathrm{a}N_\nu v_g}{2} \nonumber \\
=& \frac{\tau_\mathrm{eff}\sigma_\mathrm{a}v_g}{2\hbar V_\mathrm{eff}}\left ( \beta_\mathrm{tpa,ss} |A_s|^4 + \beta_\mathrm{tpa,pp} |A_p|^4\right.\nonumber\\
& + \beta_\mathrm{tpa,ii} |A_i|^4 + 4\beta_\mathrm{tpa,sp} |A_s|^2|A_p|^2\nonumber\\
& \left . + 4\beta_\mathrm{tpa,ip}|A_i|^2|A_p|^2 + 4\beta_\mathrm{tpa,si}|A_s|^2|A_i|^2 \right ) \label{FCAlossrate}
\end{align}
where $\sigma_\mathrm{a}$ is the free carrier absorption cross section area, including contributions from both free electrons and holes, $v_g$ is group velocity of optical modes.

\section{Minimum threshold pump power} \label{sec:Threshold pump power}
In this section we derive the minimum threshold pump power of optical parametric oscillation in a single ring cavity with traveling-wave mode. From Eq.~(\ref{eqn:Bi})--(\ref{eqn:Tp}): 
\begin{align}
T_\mathrm{p,+} = j\left(\sqrt{2\rho_\mathrm{p, ext}}\right)^{-1} \left(\rho_\mathrm{p,tot} + 8 \rho_\mathrm{i,tot}^{-1}|B_s|^2|B_p|^2\right) B_p.
\end{align}
When the input pump power is just above threshold, the OPO starts lasing, $|B_s|^2 \approx 0$, and thus 
\begin{align}
|T_{p,+}|^2 = \frac{\rho_\mathrm{p,tot}^2}{2\rho_\mathrm{p,ext}} |B_p|^2 \label{eqn:threshold}
\end{align}
The threshold pump power is the smallest pump power that can make the OPO oscillate. To minimize threshold, we can choose external coupling for pump, signal and idler to minimize the expression for pump power above (in Eq.~(\ref{eqn:threshold})). The pump power is minimized at 
\begin{align}
\rho_\mathrm{p,ext} = 1+ 2\sigma_3|B_p|^2+ \sigma_3 \rho_\mathrm{FC}^\prime |B_p|^4
\end{align}
and 
\begin{align}
P_\mathrm{th} = \left(2(1+ 2\sigma_3|B_p|^2+ \sigma_3 \rho_\mathrm{FC}^\prime |B_p|^4) |B_p|^2 \right)_\mathrm{min} \label{eqn:Pth}
\end{align}
So we need to minimize $|B_p|^2$. From Eqs.~(\ref{eqn:Bs}) and (\ref{eqn:Bi}) 
\begin{align}
2|B_p|^2 = \rho_\mathrm{s,tot} = 1+ \rho_\mathrm{s,ext}+ 4\sigma_3|B_p|^2+ \sigma_3 \rho_\mathrm{FC}^\prime |B_p|^4 \label{Bp_th}
\end{align}
By solving this quadratic equation, we have the smaller root: 
\begin{align}
|B_p|^2 = \frac{(1-2\sigma_3)-\sqrt{(1-2\sigma_3)^2-\sigma_3 \rho_\mathrm{FC}^\prime(1+\rho_\mathrm{s,ext})}}{\sigma_3 \rho_\mathrm{FC}^\prime} \label{eqn:Bp2}
\end{align}
To minimize $|B_p|^2$, we have $\rho_\mathrm{s,ext}=0$. And there is an upper limit of normalized FCA loss for OPO to oscillate: 
\begin{align}
\rho_\mathrm{FC}^\prime \le  \frac{(1-2\sigma_3)^2}{\sigma_3}
\end{align}
By putting Eq.~(\ref{eqn:Bp2}) into Eq.~(\ref{eqn:Pth}) we have the threshold pump
\begin{align}
P_\mathrm{th} = \frac{4(1-\sigma_3)}{\left((1-2\sigma_3)+\sqrt{(1-2\sigma_3)^2-\sigma_3\rho_\mathrm{FC}^\prime}\right)^2} \label{eqn:Pth2}
\end{align}
It is easy to prove that in the no FCA loss limit ($\rho_\mathrm{FC}^\prime \to 0$), the threshold pump power simplifies to Eq.~(\ref{eqn:pthpartialtpa}).  This choice of external coupling makes sense, as it corresponds to maximum parametric gain (the largest in-cavity pump light energy for given input pump power) and the smallest loss rate for the signal and idler light. 

For the case of equal pump and signal/idler coupling ($\rho_\mathrm{p,ext}=\rho_\mathrm{s,ext}=\rho_\mathrm{ext}$), Eqs.~(\ref{eqn:threshold}) and (\ref{Bp_th}) can be combined to give: 
\begin{align}
|T_{p,+}|^2 = \frac{\rho_\mathrm{p,tot}^3}{4(1-\sigma_3)\rho_\mathrm{ext}} \label{eqn:threshold_equal}
\end{align}
where $\rho_\mathrm{ext} = \rho_\mathrm{p,tot} - 1 - 2\sigma_3|B_p|^2 - \rho_\mathrm{FC}^\prime |B_p|^4$, 
$|B_p|^2 = \frac{\rho_\mathrm{s,tot}}{2} = \frac{\rho_\mathrm{p,tot}}{2-2\sigma_3}$
and then the threshold pump power can be represented by a function of $\rho_\mathrm{ext}$. This expression is complex, but can be simplified when FCA is ignorable: 
\begin{align}
P_\mathrm{th}^\prime
=& \frac{(1-\sigma_3)^2}{4(1-2\sigma_3)^3}\frac{(1+\rho_\mathrm{ext})^3}{\rho_\mathrm{ext}}|_\mathrm{min}
=& \frac{27(1-\sigma_3)^2}{16(1-2\sigma_3)^3} 
\end{align}
with $\rho_\mathrm{ext}=1/2$ at threshold.

\subsection{Understanding oscillation threshold} \label{sec:understandOscThresh}
\begin{figure}[t]
	\centering	
	\vspace{-12pt} 	 	 
	   \includegraphics[width=3.25 in]{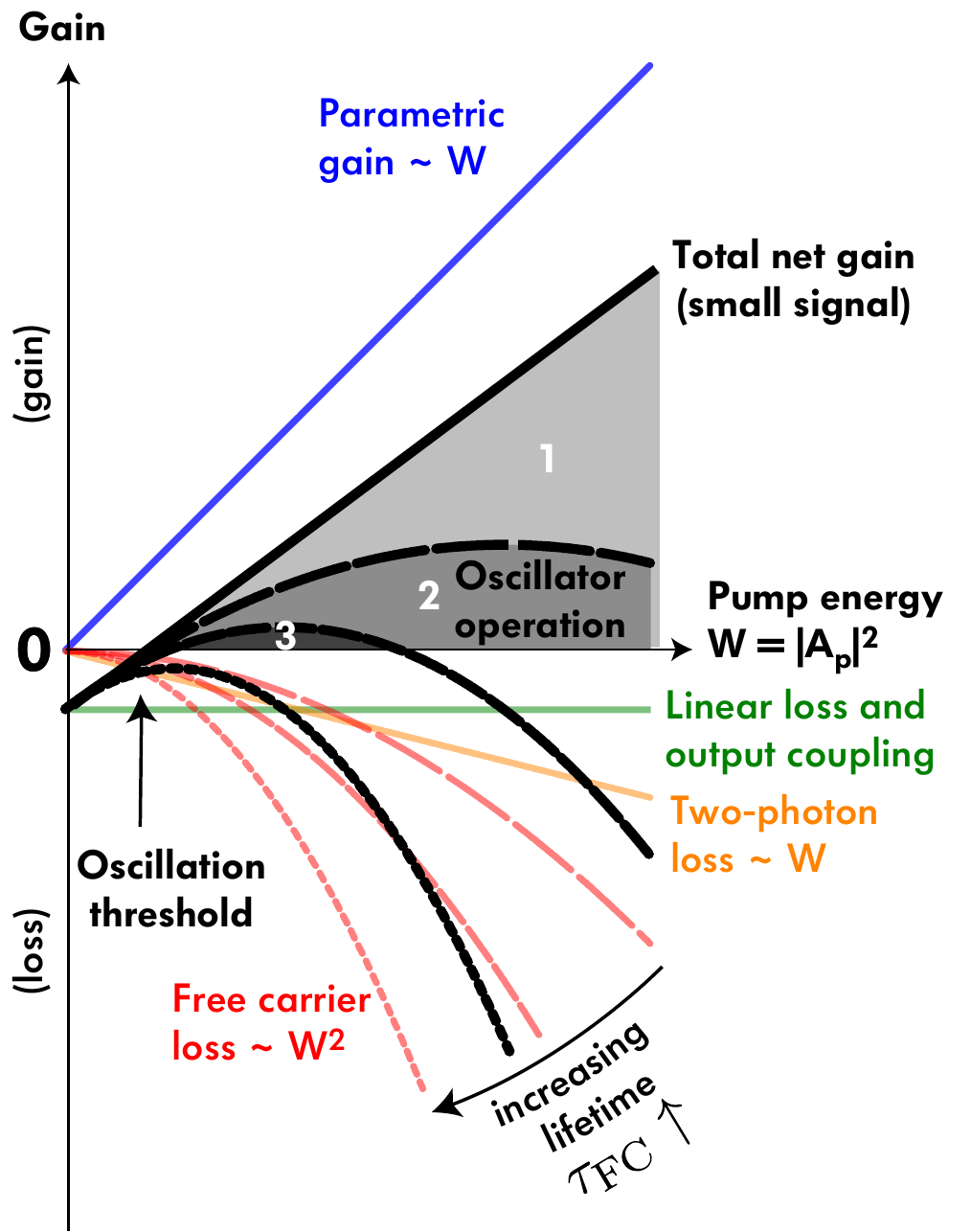}
	   \vspace{-6pt}
	\caption{Small signal gain and loss in an optical parametric oscillator based on degenerate four wave mixing.}
	\label{fig9_lossgainillustration}
\end{figure}
Here we provide a physical interpretation of the oscillation threshold when both linear and nonlinear loss are present. Fig.~\ref{fig9_lossgainillustration} shows the various terms of small-signal gain and loss for the signal resonance in an optical parametric oscillator based on degenerate four wave mixing. The linear loss rate, including material absorption, scattering loss, radiation loss and external coupling etc., is independent of the in-cavity pump energy, which roughly scales with input pump power.  The parametric gain from four wave mixing, and loss due to two photon absorption, are both proportional to pump energy. However, their scaling factors vary by a factor of $2\sigma_3$, thus it is only possible to achieve oscillation for a nonlinear material with $\sigma_3<0.5$. 

The loss due to free carrier absorption scales with the square of the pump energy, shown as parabolic curve in Fig.~\ref{fig9_lossgainillustration}. When the loss due to free carrier absorption is ignorable, the total net gain is greater than 0 in region~1. As the effective free carrier lifetime increases, the parabolic curve becomes steeper, and the region of positive net gain shrinks to region~2 and region~3. When the free carrier lifetime is above a certain limit, the total net gain is always negative, no matter how large the pump light energy is. This is consistent with the plot of oscillation threshold in Fig.~\ref{fig8_Pth}.

\section{Upper bound of FWM conversion efficiency} \label{Appendix:eta_upper_limit}
In this section we derive an upper bound of FWM conversion efficiency limited by TPA. From Eq.~(\ref{eqn:Bi})--(\ref{eqn:Tp}):
\begin{align}
T_\mathrm{p,+} &= j\left(\sqrt{2\rho_\mathrm{p, ext}}\right)^{-1} \left(\rho_\mathrm{p,tot} + 8 \rho_\mathrm{i,tot}^{-1}|B_s|^2|B_p|^2\right) B_p\nonumber\\
	   &= A + B |B_s|^2
\end{align}
Where $A\equiv j\left(\sqrt{2\rho_\mathrm{p, ext}}\right)^{-1} \rho_\mathrm{p,tot} B_p$, $B\equiv j\left(\sqrt{2\rho_\mathrm{p, ext}}\right)^{-1} 8 \rho_\mathrm{i,tot}^{-1}|B_p|^2B_p$. For the case of pump-only TPA and no FCA, the in-cavity pump light energy has a simple form (see Eq.~\ref{eqn:Ap_th}), and thus $A$ and $B$ are independent on the input pump power. The conversion efficiency is 
\begin{align}
\eta = \frac{2\rho_\mathrm{s,ext} |B_s|^2}{|T_\mathrm{p,+}|^2} = \left(\frac{2\rho_\mathrm{s, ext}}{B}\right)\frac{T_\mathrm{p,+}-A}{|T_\mathrm{p,+}|2}
\end{align}
Since the phases of A, B, $T_{p,+}$ are the same, thus 
\begin{align}
\eta = \left(\frac{2\rho_\mathrm{s,ext}}{|B|}\right)\frac{|T_\mathrm{p,+}|-|A|}{|T_\mathrm{p,+}|^2}\label{Eta}
\end{align}
It's easy to see that $\eta$ is a function only of the input pump power ($|T_\mathrm{p,+}|^2$), external coupling ($\rho_\mathrm{k,ext}$) and the nonlinear loss sine $\sigma_3$. Now we calculate the maximum conversion efficiency at fixed external coupling from $\frac{\partial \eta}{\partial |T_\mathrm{p,+}|} = 0$.  Then,
\begin{align}
|T_\mathrm{p,+}|^2 &= 4A^2 = 
\frac{2\rho_\mathrm{p,tot}^2}{\rho_\mathrm{p,ext}}|B_p|^2\nonumber
\end{align}
and substituting back into (\ref{Eta}), the maximum efficiency is
\begin{align}
\eta_\mathrm{max} 
&= \frac{\rho_\mathrm{s,ext}}{2A B}  \nonumber  \\
&=\frac{(1-2\sigma_3)^2}{2}\frac{\rho_\mathrm{s,ext}\rho_\mathrm{p,ext}\rho_\mathrm{i,tot}}{\rho_\mathrm{p,tot}(1+\rho_\mathrm{s,ext})^2}  \nonumber\\
&=\left (\frac{1}{2} - \sigma_3 \right )\frac{\rho_\mathrm{p,ext}}{\rho_\mathrm{p,tot}}\frac{\rho_\mathrm{s,ext}(1+\rho_\mathrm{i,ext})}{(1+\rho_\mathrm{s,ext})^2 }\nonumber \\ 
& < \frac{1}{2} - \sigma_3.
\end{align}
This puts an upper bound on the achievable conversion efficiency (to each of the signal and idler), as a function of the nonlinear loss sine $\sigma_3$.

\section{Analogy to laser oscillation} \label{sec:app:laseranalogy}
By combining Eq.~(\ref{eqn:dBs}) and Eq.~(\ref{eqn:Bi}) we have the lasing equation for signal light 
\begin{align}
\frac{dB_s}{d\tau} & = -\rho_\mathrm{s,tot} B_s + 4\rho_\mathrm{i,tot}^{-1}|B_p|^4 B_s \equiv -\rho_\mathrm{loss} B_s + \rho_\mathrm{gain} B_s
\end{align}
Both the loss and gain for the signal light depend on the in-cavity pump light energy. When the loss and gain term are equal, we arrive at the expression for $|B_p|^2$ in oscillation (see Eq.~(\ref{eqn:Ap_th})). At the same time, the parametric gain also depends on in-cavity signal light energy, which results in gain clamping. 
\begin{align}
\frac{dB_p}{d\tau} & = -\rho_\mathrm{p,tot} B_p -8\rho_\mathrm{i,tot}^{-1} |B_p|^2|B_s|^2B_p - j \sqrt{2\rho_\mathrm{p,ext}}T_{p,+}
\end{align} 

\bibliography{XZ_nonlinear}
\end{document}